\documentclass[aps,prb,showpacs]{revtex4}

\usepackage[intlimits]{amsmath}
\usepackage{amsfonts,graphicx}

\def\bra#1{\left\langle #1 \right|}
\def\ket#1{\left| #1 \right\rangle}
\def\aver#1{\left\langle #1 \right\rangle}

\begin{document}

\title{Rabi oscillations in semiconductor multi-wave mixing response}

\author{Mikhail Erementchouk}
\author{Michael N. Leuenberger}\email{mleuenbe@mail.ucf.edu}
\affiliation{NanoScience Technology Center and Department of Physics, University of Central
Florida, Orlando, FL 32826}

\begin{abstract}
We studied the semiconductor response with respect to high intensity
resonant excitation on short time scale when the contribution of the Fermi
statistics of the electrons and holes prevails. We studied both the single
and double pulse excitations. For the latter case we considered the time
evolution of the multi-wave mixing exciton polarization. The main
difference between the excitation by a single pulse or by two
non-collinear pulses is that the Rabi oscillations of the multi-wave
mixing response are characterized by two harmonics. Analyzing the operator
dynamics governed by the external excitation we found that there are three
invariant spin classes, which do not mix with the evolution of the system.
Two classes correspond to the bright exciton states and one contains all
dark states. We found that the dynamics of the classes is described by six
frequencies and the Rabi frequencies are only two of them (one for each
bright class). We discuss the effect of the dispersion of the electrons
and holes and the Coulomb interaction describing the semiconductor by the
semiconductor Bloch equation (SBE). We show that if initially the system
is in the ground state then the SBE preserves the invariant spin classes
thus proving absence of the dark excitons in the framework of this
description. We found that due to the mass difference between holes of
different kind additional Rabi frequencies, two of those present in the
operator dynamics, should appear in the evolution of the exciton
polarization.
\end{abstract}


\maketitle

\section{Introduction}

One of the main tools of probing the complex character of the many-body
correlations and interactions in semiconductors is the multi-wave mixing
response. The multi-wave mixing polarization in optically excited
semiconductors produces the signal in directions that are prohibited in
the linear regime because of momentum conservation, thus giving access
to the semiconductor many-body excitations. The typical example is the
four-wave mixing spectroscopy.\cite{Mukamel,Haug_Koch} Recently also wave
mixing of higher orders started to draw
attention.\cite{BOLTON:2000,CHEMLA:2001,AXT:2001} A great success in
understanding the mechanism of the formation of the nonlinear response in
general was achieved in relatively low-field limit
using the perturbation theory with respect to the external field.%
\cite{HU:1994,LINDBERG:1994,SAVASTA:1996,OSTREICH:1998,KWONG:2001,YANG:2007,EREMENTCHOUK:2007}
Perturbational description, however, is not suitable for investigating the
coherent reconstruction of the spectrum, such as in the case of Rabi
oscillations. For a qualitative analysis of such phenomena few-level
quantum models have been
used.\cite{JOFFRE:1988,CUNDIFF:1994,BONGIOVANNI:1997} This approach,
however, misses the important property of the excitations in
semiconductors. These excitations constitute a quantum field rather than a
canonical quantum mechanical system. The system supports infinitely many
states unless, of course, the spectrum has truly discrete component, that
is when the excitations are localized, such as in quantum dots. Thus a
description non-perturbative with respect to the external field must deal
with many-body aspects of the dynamics of the system.\cite{CHEMLA:2001} In
order to treat this problem a variety of methods based on derivation of
respective closed equations of motion was developed.%
\cite{SCHMITT-RINK:1988,WEGENER:1990,BINDER:1995,Haug_Koch,ROCHAT:2000,AXT:1998,TAKAYAMA:2002}
Unfortunately, the dynamical equations turn out to be very complex owing
to the Coulomb interaction, which is shown to be crucially important at
relatively long time scales.\cite{BINDER:2000,WALDMUELLER:2006} Therefore,
one has to resort to numerical calculations investigating the dynamics of
the exciton polarization. The numerical simulations proved to describe
successfully the dynamics \cite{CUNDIFF:1994,CIUTI:2000,ROCHAT:2000} but
they are difficult to apply for studying the detailed effect of different
contributions to the dynamics and its dependence on parameters of the
system. As a result, the theory of the semiconductor response suffers from
a lack of exact results obtained in controllable approximations which
would guide the respective numerical, theoretical, and experimental
studies.

At short time scales, however, one can rely on significant simplification
of the dynamics of the system due to negligible phase change during the
optical excitation. Effectively, the system follows the external field,
which is illustrated best by a simple dynamical model
\begin{equation}\label{eq:delta-source}
  \dot{P}(t) = -i\omega[|P(t)|] P(t) + E(t)
\end{equation}
with $\omega(P)$ being a real positive function. If initially the system
is at rest, $P(0)=0$, we can present the time dependence $P(t)$ in the
form
\begin{equation}\label{eq:delta_solution}
  P(t) = \int_0^t dt' \exp\left[-i \int_{t'}^t dt'' \omega[|P(t'')|]\right]
  E(t').
\end{equation}
This representation is convenient for comparing the responses of the
system with respect to the $\delta$-functional excitation $E(t) = \epsilon
\delta(t+0)$ and to the excitation with the piece-wise constant amplitude
$E(t) = \epsilon/T$, where $T$ is the duration of the forced regime. As
follows from Eq.~\eqref{eq:delta_solution}, in the case of the
$\delta$-pulse immediately after the excitation is switched off one has
$P(t) = \epsilon$. For the excitation with constant amplitude
Eq.~\eqref{eq:delta_solution} yields an estimate maximally different from
the $\delta$-pulse case in the form $P(T) = \epsilon (1-e^{-i \omega_m
T})/i \omega_m T$, where $\omega_m = \sup_{0 \leq P \leq \epsilon}
\omega(P)$. Thus if the excitation pulse is shorter than the time
scale determined by $\omega_m$, i.e. if $\omega_m T \ll 1$, then the
response with respect to the constant pulse of finite duration differs up
to quadratic terms from the $\delta$-pulse response only by the phase
factor $e^{-i \omega_m T/2}$.

An immediate application of this consideration to the evolution of the
exciton polarization in optically excited semiconductors is prevented by
the more complex character of the polarization dynamics, namely, by the
field induced coupling between the polarization and the charge densities.
In order to see the principal differences introduced by this coupling we
consider a model
\begin{equation}\label{eq:field-coupling}
  \begin{split}
  \dot{P}(t) & = -i\omega P(t) - i E(t) n(t) -i E(t), \\
  \dot{n}(t) & = i\Omega n(t) - i E(t) P(t).
  \end{split}
\end{equation}
Straightforward substitution of the $\delta$-shaped pulse into this
equation 
obviously fails since now the
amplitudes in front of the $\delta$-function in the coupling terms are
given by the quantities, $n(t)$ and $P(t)$, which experience discontinuity
at exactly the point of singularity of the $\delta$-function. Therefore,
we need to consider the case of the external pulse of finite duration. For
the pulse with piece-wise constant amplitude $E = \epsilon/T$,  as in the
previous example, Eq.~\eqref{eq:field-coupling} with the initial
conditions $P(0)=0$, $n(0)=0$  is solved for $t \leq T$ by
\begin{equation}\label{eq:field_coupling_solution}
  \begin{split}
  P(t) = & -i\frac{2\epsilon}{\Omega T} \sin(\Omega t/2)\left[\sin(\Omega t/2) -
  i\frac{\omega}{\Omega} \cos(\Omega t/2)\right], \\
 n(t) = & - \frac{\epsilon^2}{\Omega^2 T^2}\left[1- \cos (\Omega t)\right],
  \end{split}
\end{equation}
where $\Omega^2 = \omega^2/4 + E^2$. It is seen that in the limit of the
excitation pulses short compared to the typical dynamical time scale,
$\omega T \ll 1$, one has to distinguish the cases of weak and strong
excitations, $\omega T /\epsilon \gtrsim 1$ and $\omega T /\epsilon \ll 1$
respectively. In the first case up to the quadratic terms, $\Omega^2 T^2$,
the response $P(T)$ looks like the one of the system where $P$ and $n$ are
decoupled with respect to $\delta$-pulse excitation. This illustrates the
relation with the $\delta$-functional approximation popular in studying
the nonlinear optical response by using perturbational approach, which
corresponds to weak scattering.

In the limit of strong external field when $\omega T \ll 1$ while
$\epsilon \gtrsim 1$ the evolution of the system is significantly
different. Neglecting the terms quadratic in $\omega T/\epsilon$ we obtain
that the evolution of the system is similar to the one we have in the case
$\omega = 0$ (i.e. Rabi flopping) with the only difference that $P(t)$
acquires a (small) imaginary part. Applying now the similar idea for
estimating a phase shift as for the analysis of
Eq.~\eqref{eq:delta-source} we can see that the conclusion about small
perturbations vanishing with $T\to0$ holds also in the case of
$P$-dependent frequency and renormalized coupling $E \to E+\eta(|P|,
|n|)$.

These results serve as a general background for the analysis of the basic
features of the immediate response of a semiconductor excited by short
pulses of high intensity. In this regime, while taking exactly into account
the many-body effects due to the fermion nature of the electrons and holes,
one can neglect the Coulomb interaction and still be able to provide a
qualitative description. From this perspective the roles played by
statistics and interactions are clearly different. The statistics impose
instantaneous constraints on the dynamics. Interactions, in turn, need time
to develop their effect.

Explicitly the possibility to neglect
the Coulomb interaction in the semiconductor Bloch equation (SBE) was
shown for the two-band model in Ref.~\onlinecite{OSTREICH:1993}. In
Ref.~\onlinecite{WALDMUELLER:2006} the numerical investigation of the
coupled Maxwell-Bloch equations in multiple quantum well structures has
shown that one can reproduce with good accuracy a few Rabi flops at short
time scale. The case of very short excitation of very high intensity  was
studied in Ref.~\onlinecite{GOLDE:2006} from the perspective of the
carrier-wave Rabi flopping.\cite{HUGHES:1998,MUCKE:2001,MUCKE:2004} It was
demonstrated that even the spectrum of the emitted radiation is
approximated remarkably well by the free-carrier model, which corresponds
to neglecting the Coulomb interaction in the SBE. It should be noted,
however, that this extreme limit should be considered with care because of
the failure of the rotating wave
approximation.\cite{ZIOLKOWSKI:1995,HUGHES:1998}

In the present paper we extend the consideration of the semiconductor
response with respect to short intensive excitation taking into account
multiple hole states in the valence band and the formation of the
multi-wave mixing response. The structure of the paper is as follows. In
Section~\ref{sec:short-time} we consider the limiting case when the effect
of the internal dynamics can be completely neglected, which corresponds to
the cases $\Omega = 0$ in the examples considered above. We obtain classes
of state related to the spin selection rules and related to these classes
the spectrum of the Rabi frequencies. In Section~\ref{sec:internal} we
discuss the effect of the internal dynamics and the Coulomb interaction.


\section{Rabi oscillations}
\label{sec:short-time}

We consider the dynamics of excitation of a semiconductor quantum well
during the action of an ultrashort pulse of high intensity so that the
period of the Rabi oscillations is smaller than the typical time scales
determined by the internal dynamics. We consider the question of validity
of this approximation in Section~\ref{sec:internal}. We assume that the
pulse is tuned to resonance with the exciton levels lying below the band
edge and that the envelope of the pulse is constant while the field is on.
By means of a canonical transformation the harmonic time dependence of the
excitation field can be excluded, (see e.g.
Ref.~\onlinecite{LINDBERG:1992}) so that in the rotating wave
approximation the transformed Hamiltonian becomes time-independent. For
the  description of the semiconductor dynamics one can use the SBE
accounting for degenerate valence
bands.\cite{BINDER:1995,BINDER:1997,BINDER:2000} Having the dynamical
equations at hand one can work out the short time approximation similarly
to the one shown in the Introduction. However, in the short time
approximation, it is more convenient to study directly the exciton
polarization. In the coherent regime it is defined as
  $P_\mu = \langle \psi(t)|B_\mu | \psi(t) \rangle$,
where $|\psi(t) \rangle$ is the state of the semiconductor, $\mu$ denotes
the whole set of relevant quantum numbers describing the specific exciton
(bound or unbound) state and $B_\mu$ is the exciton annihilation operator
defined as $\langle 0| B_\mu = \langle \mu|$. In terms of the electron and
hole operators the exciton operator $B_\mu$ is presented as
\begin{equation}\label{eq:exciton_operator_elementary}
  B_\mu = \int d\mathbf{x} d\mathbf{y} \,\phi^*_\mu(\mathbf{x},\mathbf{y})
  c_{s_\mu}(\mathbf{x})
  v_{\sigma_\mu}(\mathbf{y}),
\end{equation}
where $\phi_\mu(\mathbf{x},\mathbf{y})$ is the exciton wave functions,
$v_{\sigma_\mu}(\mathbf{y})$ destroys a hole in the valence band with the
spin state $\sigma_\mu$ at the point with the coordinate $\mathbf{y}$ and
$c_{s_\mu}(\mathbf{x})$ destroys an electron in the conduction band with
the spin state $s_\mu$ at point $\mathbf{x}$.

Neglecting the Coulomb interaction and the effect of slow internal
dynamics determined by small detuning $\omega_\mu$ means the semiconductor
Hamiltonian is approximated by the Hamiltonian of light-matter interaction
$H_e$. In the rotating wave and dipole approximations the interaction
Hamiltonian has the form \cite{KIRA:1999,KHITROVA:1999,Haug_Koch}
\begin{equation}\label{eq:perturbation_dipole}
  H_e = 
  \sum_\mu \left[B_\mu \epsilon^*_\mu(t) + B^\dagger_\mu
  \epsilon_\mu(t)\right],
\end{equation}
where $\epsilon_\mu(t)$ is a piece-wise constant function of time, it
takes the values $0$ when the field is off and $\epsilon_\mu
 = \int d\mathbf{x}\, \mathbf{d}_\mu
\cdot\mathbf{E}(\mathbf{x}) \phi^*_\mu(\mathbf{x},\mathbf{x})$ with
$\mathbf{E}(\mathbf{x})$ being the envelope of the external field when the
excitation is on. Here we have introduced $\mathbf{d}_\mu$ the respective
matrix elements of the dipole moment, which are assumed to be independent
on both the electromagnetic field momentum and its angle of
incidence.\cite{KIRA:1999} Thus, in this approximation one has $|\psi(t)
\rangle = \exp(-i H_e t)|0 \rangle$ (throughout the paper we write
formulas using units with $\hbar = 1$) and accordingly
\begin{equation}\label{eq:P_t_series_presentation}
  P_\mu(t) = \sum_n \frac{(it)^n}{n!}\aver{[H_e,[H_e,\ldots [H_e, B_\mu]\ldots]]},
\end{equation}
where $\langle \ldots \rangle \equiv \langle 0| \ldots | 0 \rangle$. We
calculate this series using the observation that commuting $[B_\mu,
B^\dagger_\nu]$ with $H_e$ we arrive at an operator which can be
presented in a form similar to Eq.~(\ref{eq:exciton_operator_elementary}).
We employ this observation introducing operators
\begin{equation}\label{eq:M_op_defs}
  M^{(n)} = \sum_{\sigma, s}\int d\mathbf{x} d\mathbf{y}\,\left[
  U^{(n)}_{\sigma, s}(\mathbf{x},\mathbf{y}) v^\dagger_\sigma(\mathbf{x})c^\dagger_s(\mathbf{y})
+ D^{(n)}_{\sigma, s}(\mathbf{x},\mathbf{y}) c_s(\mathbf{y})v_\sigma(\mathbf{x})
  \right],
\end{equation}
which are related to each other through  repetitive commutation with
$H_e$, namely,
  $M^{(n+1)} = [H_e,[H_e,  M^{(n)}]]$.
With the ``initial condition" $M^{(0)} = B_\mu$ the operators $M^{(n)}$
represent the \textit{even} terms in
series~(\ref{eq:P_t_series_presentation}). These terms themselves do not
contribute into $P_\mu$ since $\aver{M^{(n)}} = 0$, which in turn follows
from $c |0\rangle = 0$ and $\langle 0|c^\dagger  = 0$. After commuting
with $H$, however, they produce $c$-number terms $A^{(n)} = \aver{[H,
M^{(n)}]}$ resulting in
\begin{equation}\label{eq:P_mu_sum_A}
  P_\mu(t) = i\sum_{n=0}^\infty (-1)^{n}\frac{t^{2n+1}}{(2n+1)!} A^{(n)}.
\end{equation}

In order to simplify the expressions we assume below that the quantum well
can be approximated by a 2D plane. In this case the exciton states are
characterized by the spin states of the electron and hole constituting the
exciton, the center of mass momentum in the plane of the well,
$\mathbf{K}$, and other quantum numbers, $n_\mu$, so
that $\ket{\mu} = \ket{\sigma_\mu, s_\mu, \mathbf{K}_\mu, n_\mu}$.

Using the relation between $M^{(n+1)}$ and $M^{(n)}$ we find
\begin{equation}\label{eq:kernels_relation}
\begin{split}
U^{(n+1)}_{\sigma, s}(\mathbf{x},\mathbf{y}) = \sum_{\sigma', s'}
 & \left[
 U^{(n)}_{\sigma', s}(\mathbf{x},\mathbf{y})\mathcal{E}_{\sigma, s'}(\mathbf{x})
 \mathcal{E}^*_{\sigma', s'}(\mathbf{x}) \right. \\
 & \left. \qquad + U^{(n)}_{\sigma, s'}(\mathbf{x},\mathbf{y})\mathcal{E}_{\sigma', s}(\mathbf{y})
 \mathcal{E}^*_{\sigma', s'}(\mathbf{y}) 
 -2 D^{(n)}_{\sigma', s'}(\mathbf{y},\mathbf{x})\mathcal{E}_{\sigma, s'}(\mathbf{x})
 \mathcal{E}_{\sigma', s}(\mathbf{y})
 \right] \\
D^{(n+1)}_{\sigma, s}(\mathbf{x},\mathbf{y})  = \sum_{\sigma', s'}
 & \left[
 D^{(n)}_{\sigma', s}(\mathbf{x},\mathbf{y})\mathcal{E}^*_{\sigma, s'}(\mathbf{x})
 \mathcal{E}_{\sigma', s'}(\mathbf{x})\right. \\
 & \left. \qquad +D^{(n)}_{\sigma, s'}(\mathbf{x},\mathbf{y})\mathcal{E}^*_{\sigma', s}(\mathbf{y})
 \mathcal{E}_{\sigma', s'}(\mathbf{y}) 
 -2 U^{(n)}_{\sigma', s'}(\mathbf{y},\mathbf{x})\mathcal{E}^*_{\sigma, s'}(\mathbf{x})
 \mathcal{E}^*_{\sigma', s}(\mathbf{y})
 \right],
\end{split}
\end{equation}
where $\mathcal{E}_{\sigma, s}(\mathbf{x}) = d_{\sigma, s}
E_{s-\sigma}(\mathbf{x})$. In terms of the kernels of the operators
$M^{(n)}$ the coefficients $A^{(n)}$ are expressed as $A^{(n)} = \int
d\mathbf{x} \,A^{(n)}(\mathbf{x})$ with
\begin{equation}\label{eq:A_kernels}
    A^{(n)}(\mathbf{x}) = \sum_{\sigma, s} \left[
 \mathcal{E}^*_{\sigma, s}(\mathbf{x})U^{(n)}_{\sigma, s}(\mathbf{x},\mathbf{x})
 - \mathcal{E}_{\sigma, s}(\mathbf{x})D^{(n)}_{\sigma, s}(\mathbf{x},\mathbf{x})
  \right].
\end{equation}

Some general results can be obtained directly from
Eqs.~(\ref{eq:kernels_relation}) and (\ref{eq:A_kernels}). First, the
polarization of dark excitons (states with helicity $0$ and $\pm2$) is zero.
Second, if the excitation pulse is circularly polarized then summation over
the spin indices in Eqs.~(\ref{eq:kernels_relation})
and~(\ref{eq:A_kernels}) reduces to the single terms with the electron and
hole spins determined by $s_\mu$ and $\sigma_\mu$, respectively. If the
excitation pulse is linearly polarized then only the electron spin is fixed
to $s_\mu$ and as a result different electron-hole states get
coupled.\cite{BINDER:1995} The summation over the hole spins, as will be
explicitly demonstrated below, to be limited to such values that meet the
condition $\sigma - \sigma_\mu \in \{2, 0, -2\}$. For example, if
$\sigma_\mu = 3/2$ (heavy holes) then $\sigma$ can take values $3/2$ and
$-1/2$ (light hole).

It follows from Eqs.~(\ref{eq:kernels_relation}) and (\ref{eq:A_kernels})
that
  $A^{(n+1)}(\mathbf{x}) = \Omega^2_\mu(\mathbf{x}) A^{(n)}(\mathbf{x})$,
where
\begin{equation}\label{eq:RabiF_x}
 \Omega^2_\mu(\mathbf{x}) =4
 \sum_{\sigma = \sigma_\mu, \bar{\sigma}_\mu} \left|\mathcal{E}_{\sigma, s_\mu}(\mathbf{x})\right|^2
\end{equation}
with $\bar{\sigma}_\mu$ being the second value satisfying the spin
selection rule discussed above. Applying these results to
Eq.~(\ref{eq:P_mu_sum_A}) we find the general representation, which is
valid for an excitation pulse with an arbitrary spatial profile in the
plane of the quantum well,
\begin{equation}\label{eq:p_mu}
  P_\mu(t) = -i \int d\mathbf{x}\, \frac{1}{\Omega_\mu(\mathbf{x})}
  \mathcal{E}_{\mu}(\mathbf{x}) \phi^*_\mu(\mathbf{x},\mathbf{x}) \sin\left[\Omega_\mu(\mathbf{x}) t\right].
\end{equation}
This result can be generalized for the more general case when the envelope
amplitude of the external field is not a piece-wise constant and can be
presented as $\mathbf{E}(\mathbf{x}, t) = f(t)\mathbf{E}(\mathbf{x})$. We
note that the Hamiltonians $H'_e(t) = f(t) H_e$ taken at different instants
$t_1$ and $t_2$ commute with each other. Therefore, the line of arguments
used above can be repeated with the substitution
\begin{equation}\label{eq:time_parameter}
  t \to \int_0^t dt'\,f(t').
\end{equation}
It can be easily seen that the expression for the exciton polarization
$P_\mu(t)$ remains essentially the same
\begin{equation}\label{eq:p_mu_smooth}
  P_\mu(t) = -i \int d\mathbf{x}\, \frac{1}{\Omega_\mu(\mathbf{x}, t)}
  \mathcal{E}_{\mu}(\mathbf{x}, t) \phi^*_\mu(\mathbf{x},\mathbf{x}) \sin\left[\varpi_\mu(\mathbf{x},t)\right],
\end{equation}
where $\mathcal{E}_{\mu}(\mathbf{x}, t)$ and $\Omega_\mu(\mathbf{x}, t)$ are
defined by the same expressions but with the time-dependent envelope
function of the external field, i.e. they differ by the factor $f(t)$. The
phases
\begin{equation}\label{eq:effective_areas}
 \varpi_\mu(\mathbf{x},t) = \int_0^t dt'\, \Omega_\mu(\mathbf{x},t)
\end{equation}
have the meaning of the pulse areas. Since such generalization does not
bring new physics but complicates the discussion of the time dependence in
what follows we will consider only the case of piece-wise constant amplitude
of the external excitation.

\subsection{Single-pulse response}

First we briefly discuss the case when the excitation is a single plane wave
$\mathbf{E}(\mathbf{x}) = (E_+ \widehat{\mathbf{e}}_+ + E_-
\widehat{\mathbf{e}}_-)e^{i\mathbf{K}_0\cdot \mathbf{x}}$. In this case
$\Omega_\mu(\mathbf{x})$ does not depend on the coordinate and the
integration in Eq.~\eqref{eq:p_mu} can be easily performed taking into
account that $\phi^*_\mu(\mathbf{x},\mathbf{y}) = e^{-I \mathbf{K}_\mu\cdot
\mathbf{R}_\mu} \phi^*_\mu(\mathbf{x} -\mathbf{y})$, where $\mathbf{R}_\mu$
is the coordinate of the center of mass of the exciton and
$\phi^*_\mu(\mathbf{x} -\mathbf{y})$ is its relative wave function. For
example, for $(h+, 1s)$ exciton polarization we obtain
\begin{equation}\label{eq:plane_wave_excitation_h}
  P_{h+, \mathbf{K}}(t) = -i \delta(\mathbf{K} - \mathbf{K}_0) d_{h+} E_+
  \phi^*_{h+}(0) \frac{1}{\Omega_{h+}} \sin(\Omega_{h+} t),
\end{equation}
where $\phi^*_{h+}(0)$ is the value of the exciton wave function at the
origin, which in the 2D approximation is $\phi^*_{h+}(0) =
r_{h+}^{-1}\sqrt{2/\pi}$ with $r_{h+}$ being the respective exciton Bohr
radius, and
\begin{equation}\label{eq:RabiF_h}
  \Omega_{h+} = 2\sqrt{\left|d_{h+}E_+\right|^2 + \left|d_{l-}E_-\right|^2}.
\end{equation}
Equation (\ref{eq:plane_wave_excitation_h}) presents the Rabi oscillations
of the exciton polarization. It is interesting to note that this result
explicitly shows the statistical origin of the Rabi oscillations. Following
the same line of arguments one can show that the oscillations would absent and
the exciton polarization would increase monotonously with time if the
exciton operators had obeyed bosonic commutation relations or if the
electrons and holes were bosons. Indeed, let, for example, the exciton
operators be bosonic then all terms in
Eq.~\eqref{eq:P_t_series_presentation} with $n > 1$ turn to $0$ leaving
$P_\mu \propto t$.

\begin{figure}
  \includegraphics[width=3in]{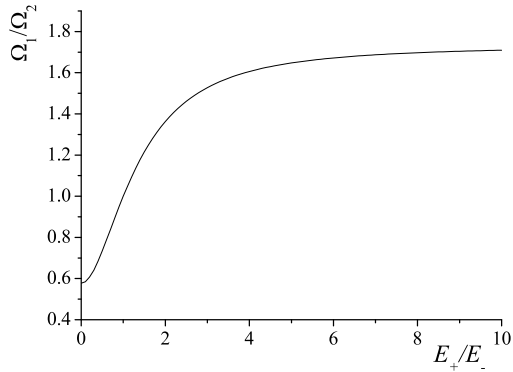}
  \caption{The ratio of two Rabi frequencies $\Omega_{h+}/\Omega_{h-}$
  as a function of ellipticity of the excitation pulse $E_+/E_-$. The ratio varies
  from $|d_{l}/d_{h}|$ to $|d_{h}/d_{l}|$. For calculations the ratio
  of heavy- and light-hole dipole moments is taken $d_h/d_l = \sqrt{3}$.
  }\label{fig:Rabi_ratio}
\end{figure}

Because of the coupling between different exciton states the evolution of
the exciton polarization is characterized by two Rabi frequencies,
$\Omega_{h+} =\Omega_{l-}$ and $\Omega_{h-} =\Omega_{l+} =
2\sqrt{\left|d_{h-}E_-\right|^2 + \left|d_{l+}E_+\right|^2}$. The ratio
between these frequencies is shown in Fig.~\ref{fig:Rabi_ratio} as a
function of $E_+/E_-$, where for calculations we have
taken\cite{Yu_Cardona} $d_h/d_l = \sqrt{3}$.

The effect of sharing the Rabi frequency leads to the same time dependence
of the heavy- and light-hole contributions into the signal with fixed
helicity if the excitation is linearly polarized. If, however, the
external field has elliptic polarization these contributions oscillate
with different frequencies leading to beatings and to a nontrivial time
dependence of the polarization state of the signal.

\subsection{The operator equations of motion}

Before we proceed we would like to discuss in details the spin selection
rules noticed above from the perspective of the operator equations of
motion. For this we consider an operator of the form
\begin{equation}\label{eq:F_operator}
\begin{split}
  \widehat{F}(t;\mathbf{x},\mathbf{y}) = &
 f_{\sigma_1, \sigma_2}(t)v^\dagger_{\sigma_1}(\mathbf{x})
  v_{\sigma_2}(\mathbf{y}) + g_{s_1, s_2}(t)c^\dagger_{s_1}(\mathbf{x})c_{s_2}(\mathbf{y}) + \\
  &\pi^{(1)}_{\sigma, s}(t)v^\dagger_{\sigma}(\mathbf{x})c^\dagger_{s}(\mathbf{y})
  + \pi^{(2)}_{\sigma, s}(t)c_{s}(\mathbf{y})v_{\sigma}(\mathbf{x})+ A(t)\delta(\mathbf{x}-\mathbf{y}),
\end{split}
\end{equation}
where the sum is taken over all spin indices. The amplitudes in this
expression are chosen in such a way that
$\widehat{F}(t;\mathbf{x},\mathbf{y})$ is the Heisenberg representation of
the operator $\widehat{F}(0;\mathbf{x},\mathbf{y})$ defined by the
light-matter Hamiltonian
\begin{equation}\label{eq:F_Heizenberg}
 \widehat{F}(t;\mathbf{x},\mathbf{y}) = e^{i H_e t}\widehat{F}(0;\mathbf{x},\mathbf{y})e^{-i H_e t}.
\end{equation}
The last term in Eq.~\eqref{eq:F_operator} determines the average value
$\aver{\widehat{F}(t;\mathbf{x},\mathbf{y})}$ and for the case of the
exciton polarization was considered above. For the purpose of our
discussion it suffices to consider the dynamical equations for the
time-dependent amplitudes $g_{\sigma_1, \sigma_2}(t)$ and so on for the
case of $\mathbf{x}=\mathbf{y}$ and spatially independent external field
$\mathcal{E}_{\sigma,s}$. From the Heisenberg equation of motion $-i
\dot{\widehat{F}} = [H_e,\widehat{F}]$ we find (see
Appendix~\ref{app:external_field_algebra})
\begin{equation}\label{eq:amplitudes_dynamics}
  \begin{split}
  \dot{f}_{s_1,s_2} & = i \mathcal{E}_{\sigma',s_1} \pi^{(2)}_{\sigma',s_2} -
  i \pi^{(1)}_{\sigma',s_1} \mathcal{E}^*_{\sigma',s_2}, \\
 \dot{g}_{\sigma_1,\sigma_2} & = i \mathcal{E}_{\sigma_1,s'} \pi^{(2)}_{\sigma_2,s'} -
  i \pi^{(1)}_{\sigma_1,s'}\mathcal{E}^*_{\sigma_2,s'}, \\
 \dot{\pi}^{(1)}_{\sigma,s} & = -i \mathcal{E}_{\sigma,s'} f_{s,s'}
  - i \mathcal{E}_{\sigma',s} g_{\sigma,\sigma'}, \\
 \dot{\pi}^{(2)}_{\sigma,s} & = i \mathcal{E}^*_{\sigma,s'} f_{s',s}
  + i\mathcal{E}^*_{\sigma',s} g_{\sigma',\sigma},
  \end{split}
\end{equation}
where the summation over the dashed spin variables is implied. With the
respective initial conditions these equations define the Heisenberg
representation of the operator $\widehat{F}(0;\mathbf{x},\mathbf{y})$. For
example, if all amplitudes but $\pi^{(1)}_{\sigma,s}$ are initially zero,
then this would be the representation of the \emph{interband} polarization
creation operator and so on. It should be noted that no restrictions on
the amplitudes being ``bright" or ``dark" are imposed in
Eq.~\eqref{eq:amplitudes_dynamics}. Thus this is a system with respect to
$36$ unknowns and, counting degeneracy, has as many characteristic
frequencies. The actual number of different frequencies, however, turns
out to be much smaller.


Taking the derivative of the last pair of equations with respect to time
we obtain the closed system
\begin{equation}\label{eq:polarization_amplitudes}
  \begin{split}
  \ddot{\pi}^{(1)}_{\sigma,s} & =-\mathcal{E}_{\sigma,s'}\mathcal{E}^*_{\sigma',s'} \pi^{(1)}_{\sigma',s}
  - \pi^{(1)}_{\sigma,s'}\mathcal{E}^*_{\sigma',s'}\mathcal{E}_{\sigma',s} 
  + 2 \mathcal{E}_{\sigma,s'}\pi^{(2)}_{\sigma',s'}\mathcal{E}_{\sigma',s}, \\
  \ddot{\pi}^{(2)}_{\sigma,s} & =-\mathcal{E}^*_{\sigma,s'}\mathcal{E}_{\sigma',s'}\pi^{(2)}_{\sigma',s}
  - \pi^{(2)}_{\sigma,s'}\mathcal{E}^*_{\sigma',s'}\mathcal{E}_{\sigma',s}
  + 2 \mathcal{E}^*_{\sigma,s'}\pi^{(1)}_{\sigma',s'}\mathcal{E}^*_{\sigma',s},
  \end{split}
\end{equation}
where as as well the summation over dashed spin indices is performed.
Taking the complex conjugation of these equations we obtain the same
system but with respect to $({\pi^{(2)}}^*, {\pi^{(1)}}^*)$.
Because of this symmetry the solutions of
Eq.~\eqref{eq:polarization_amplitudes} are divided into two classes
$\pi^{(2)} = {\pi^{(1)}}^*$ and $\pi^{(2)} = -{\pi^{(1)}}^*$, which
correspond to $A$ and $B$ irreducible representations of the group $Z_2$,
respectively. Additional simplifications come from the structure of the
coupling induced by the external field. Having in mind successive
application for the dynamics determined by the semiconductor Bloch
equation we consider the additional symmetries on a general ground. Due to
the electric dipole selection rules only four elements of
$\mathcal{E}_{\sigma,s}$ are non-zero,
\begin{equation}\label{eq:general_external_field}
\mathcal{E}_{\sigma,s} =
\begin{pmatrix}
 \mathcal{E}_{h-} & 0 \\
 0 & \mathcal{E}_{l-} \\
 \mathcal{E}_{l+} & 0 \\
  0 &  \mathcal{E}_{h+}
\end{pmatrix},
\end{equation}
where the columns are enumerated by the electron spin projections $1/2$,
$-1/2$ and rows denote different hole spins $3/2, \ldots, -3/2$. We denote
by $\mathcal{V}_{\mathcal{E}}$ the vector space containing the elements of
form~\eqref{eq:general_external_field} but without any relation between
the nonzero elements, that is the $4$-dimensional subspace of the
interband polarizations. For two elements $u$, $v \in
\mathcal{V}_{\mathcal{E}}$ we consider $f = v^T u$, where the composition
is the usual matrix multiplication of two (rectangular) matrices
\begin{equation}\label{eq:V-e_definition}
  f_{s_1,s_2} = \sum_{\sigma'} v_{\sigma',s_1} u_{\sigma',s_2}.
\end{equation}
The space spanned by such elements will be called $\mathcal{V}_e$. This is
a $2$-dimensional subspace of the $4$-dimensional space of the electron
densities. Using Eq.~\eqref{eq:general_external_field} in
\eqref{eq:V-e_definition} one can see that in the basis of the
electron-hole spins $\mathcal{V}_e$ consists of diagonal $2\times2$
matrices. Finally, for the elements $u$ and $v$ we consider $g = v u^T$,
which are the matrices with the matrix elements
\begin{equation}\label{eq:V-h_definition}
  g_{\sigma_1,\sigma_2} = \sum_{s'} v_{\sigma_,s'} u_{\sigma_2,s'}.
\end{equation}
We will call the respective linear vector space $\mathcal{V}_h$, which, as
can be easily checked, is $8$-dimensional subspace of the $16$-dimensional
space of the hole-hole correlation functions.

The elements of $\mathcal{V}_e$ and $\mathcal{V}_h$ can be considered as
operators acting on the electron and the hole spin variables,
respectively. The important properties of these operators is that their
action is reducible, namely, $\mathcal{V}_e$ leaves the spin state intact
and $\mathcal{V}_h$ mixes only states within the groups $\lbrace 3/2,-1/2
\rbrace$ and $\lbrace -3/2,1/2 \rbrace$ but not from the different groups.

Using these results we obtain the modes of
system~\eqref{eq:polarization_amplitudes}. These modes are naturally
classified by the invariant spin classes, which do not mix during the
evolution, and the irreducible representations of the group $Z_2$.
Inspecting Eqs.~\eqref{eq:polarization_amplitudes} one can see that the
spin classes in the spin space of the electron-hole polarizations are $C_1
= \{(-3/2, -1/2), (1/2,-1/2)\}$ or $\{h+,l-\}$, $C_2= \{(3/2, 1/2),
(-1/2,1/2)\}$ and $C_3=\{(3/2,-1/2),(1/2,1/2),(-1/2,-1/2),(-3/2, 1/2)\}$.
The first two classes correspond to the bright exciton operators, the last
class
contains the dark excitons.

\subsubsection{The bright exciton classes $C_{1,2}$}

In each class both states have the same projection of the electron spin,
$s=1/2$ for $C_1$ and $s=-1/2$ for $C_2$, while the hole spin can take two
values. If we need to distinguish these values we will denote them
$\sigma$ and $\bar{\sigma}$. In order to find the modes one can present
$\pi^{(1)}_{\sigma,s}(t) = \pi^{(1)}_{\sigma,s} \sin(\omega t)$ in
Eq.~\eqref{eq:polarization_amplitudes}. Studying the resulting system of
homogeneous equations one finds the frequencies and (non-normalized)
eigen-vectors of the modes
\begin{equation}\label{eq:c12_states}
  \begin{split}
  \omega^2 & = 4\beta_s^2, \quad \pi^{(1)}_{\sigma,s} = \mathcal{E}_{\sigma,s}, \quad (B)\\
  \omega^2 & = \beta_s^2, \quad \pi^{(1)}_{\sigma,s} = \frac{1}{\mathcal{E}_{\sigma,s}},
  \quad \pi^{(1)}_{\bar{\sigma},s} = -\frac{1}{\mathcal{E}_{\bar{\sigma},s}}, \quad (A, B)\\
  \omega^2 & = 0, \quad \pi^{(1)}_{\sigma,s} = \mathcal{E}_{\sigma,s}, \quad (A),
  \end{split}
\end{equation}
where $\beta_{s}^2 = |\mathcal{E}_{\sigma,s}|^2 +
|\mathcal{E}_{\bar{\sigma},s}|^2$ ( $4\beta^2_s$ coincides with the
frequency defined in Eq.~\eqref{eq:RabiF_x}) and $A$ and $B$ denote the
even and odd irreducible representations, respectively. Using these
solutions in the first two equations of
Eqs.~\eqref{eq:amplitudes_dynamics} we find the electron and hole
densities coupled to the respective interband polarization operators. One
can immediately see that these densities belong to $\mathcal{V}_e$ and
$\mathcal{V}_h$, respectively.

As immediately follows from Eqs.~\eqref{eq:A_kernels} and
\eqref{eq:amplitudes_kernels_dynamics} only $B$-solutions can contribute
into the average value of the exciton polarization, since for the
symmetric solutions both terms under the integral in
Eq.~\eqref{eq:amplitudes_kernels_dynamics} cancel each other. Moreover,
for the $B$-solutions with the frequencies $\omega^2 = \beta_s^2$ the sum
$\pi^{(1)}_{\sigma,s}\mathcal{E}_{\sigma,s} +
\pi^{(1)}_{\bar{\sigma},s}\mathcal{E}_{\bar{\sigma},s}$ vanishes, which
leaves only solution with $\omega^2 = 4\beta_s^2$ in accordance with
Eq.~\eqref{eq:p_mu}.

\subsubsection{The dark exciton class $C_{3}$}

The dark exciton operators are specified by the amplitudes in the
complement to $\mathcal{V}_{\mathcal{E}}$. The states are naturally
enumerated by the hole spin state $\sigma$ and for each amplitude
$\pi_{\sigma,s}$ there is a correspondent component of the excitation
field obtained by inversion of the electron spin
$\mathcal{E}_{\sigma,\bar{s}}$ with $\bar{s}\equiv - s$. The frequencies
and non-zero components of different modes are found to be
\begin{equation}\label{eq:c3_states}
  \begin{split}
  \omega^2 & = \beta_{1/2}^2, \quad \pi^{(1)}_{3/2,-1/2} = -\frac{1}{\mathcal{E}_{3/2,1/2}},
  \quad
  \pi^{(1)}_{-1/2,-1/2} = \frac{1}{\mathcal{E}_{-1/2,1/2}}, \quad (A,B)\\
  \omega^2 & = \beta_{-1/2}^2, \quad
  \pi^{(1)}_{1/2,1/2} = -\frac{1}{\mathcal{E}_{1/2,-1/2}}, \quad
  \pi^{(1)}_{-3/2,1/2} = \frac{1}{\mathcal{E}_{-3/2,-1/2}}, \quad (A,B)\\
  \omega^2 & = \left(\beta_{1/2}\pm \beta_{-1/2}\right)^2, \quad \pi^{(1)}_{3/2,-1/2} = \pi^{(1)}_{-1/2,-1/2} = \mp \beta_{1/2},
  \quad \pi^{(1)}_{-3/2,1/2} = \pi^{(1)}_{1/2,1/2} = \mp \beta_{-1/2}, \quad (A), \\
  \omega^2 & = \left(\beta_{1/2}\pm \beta_{-1/2}\right)^2, \quad \pi^{(1)}_{3/2,-1/2} = \pi^{(1)}_{-1/2,-1/2} = \pm \beta_{1/2},
  \quad \pi^{(1)}_{-3/2,1/2} = \pi^{(1)}_{1/2,1/2} = \pm \beta_{-1/2}. \quad (B)
  \end{split}
\end{equation}
Since for each $\pi_{\sigma,s}$ in this class the respective components of
the external field are zero, i.e. $\mathcal{E}_{\sigma,s}=0$, the dark
exciton amplitudes, according to
Eq.~\eqref{eq:amplitudes_kernels_dynamics}, do not contribute to
$\dot{A}(t)$. Thus the average value of the dark exciton polarization
equals to its initial value $\widehat{F}_{\sigma,s}(t) =
\widehat{F}_{\sigma,s}(0)$, where $\widehat{F}_{\sigma,s}(0)=
v^\dagger_\sigma c^\dagger_s$ with $\sigma$ and $s$ from $C_3$. If the
initial state of the system is vacuum then
$\langle\widehat{F}_{\sigma,s}(t)\rangle = 0$ confirming the result about
absent polarization of dark excitons.

\subsubsection{The operator dynamics}

The frequencies found $\omega^2 = 4 \beta_s^2,\beta_s^2, (\beta_s \pm
\beta_{\bar{s}})^2$ define all the frequencies introduced into the
dynamics due to the interaction with the external field as defined by
Eq.~\eqref{eq:amplitudes_dynamics}. The remaining $8$ frequencies can be
shown to be $0$. Four solutions (two for each class $C_1$ and $C_2$)
corresponding to the zero frequency are presented in
Eqs.~\eqref{eq:c12_states} and the remaining $4$ can be found from the
first two equations of Eqs.~\eqref{eq:amplitudes_dynamics}. Since
Eqs.~\eqref{eq:amplitudes_dynamics} have the form of a Schr\"{o}dinger
equation with a hermitian Hamiltonian the solutions corresponding to
$\omega = 0$ are not secular solutions but rather invariants, that is they
correspond to operators, which are invariant with respect to action of the
external field. For example, as follows from Eqs.~\eqref{eq:c12_states},
for the exciton polarization such operators are
$\sum_{\sigma,s}\mathcal{E}_{\sigma,s} v^\dagger_\sigma c^\dagger_s +
h.c.$, where the spin summation is restricted to a particular class
$C_{1}$ or $C_{2}$.

In the short-time limit only frequencies $4\beta_s^2$ contribute to the
Rabi oscillations of the exciton polarization. However, as will be
demonstrated later the slow dynamics leads also to admixture of other
frequencies.

\subsection{Multi-wave mixing response}

A more complex situation arises when the excitation pulse consists of two
plane waves with non-collinear wave-vectors, $\mathbf{E}(\mathbf{K}) =
\mathbf{E}^{(1)} \delta(\mathbf{K} - \mathbf{K}_1) + \mathbf{E}^{(2)}
\delta(\mathbf{K} - \mathbf{K}_2)$. This corresponds to the standard
problem of excitation by two pulses with zero delay between them.
Factoring out the exponential factor corresponding to the first pulse we
can rewrite Eq.~(\ref{eq:p_mu}) as
\begin{equation}\label{eq:integral_copied}
  \int d\mathbf{x} e^{i\mathbf{k}\cdot\mathbf{x}}F(\mathbf{x}),
\end{equation}
where $F(\mathbf{x})$ is a periodic function $F(\mathbf{x}+\Delta
\mathbf{K}/(2\pi \Delta K^2)) = F(\mathbf{x})$ with $\Delta \mathbf{K} =
\mathbf{K}_2 - \mathbf{K}_1$. Thus, integration over $\mathbf{x}$ yields
the exciton polarization in the form
\begin{equation}\label{eq:polarization_sum_MWW}
  P_\mu(t) = \sum_{m=-\infty}^\infty P_\mu^{(m)}(t)
  \delta(\mathbf{K}_\mu - \mathbf{K}_1 - m \Delta \mathbf{K}),
\end{equation}
where $P_\mu^{(m)}(t)$ are the amplitudes of multi-wave mixing (MWM)
polarizations. The amplitudes $P_\mu^{(0)}(t)$ and $P_\mu^{(1)}(t)$
correspond to the polarization along the directions of linear response,
$\mathbf{K}_1$ and $\mathbf{K}_2$, respectively. For $m=-1,2$ one has
four-wave mixing polarization, $m=-2,3$ correspond to six-wave mixing and
so on. Performing the Fourier transform of the periodic factor
$F(\mathbf{x})$ we derive the integral representation for the amplitudes
of MWM polarization
\begin{equation}\label{eq:n_amplitudes_mwm}
  P_\mu^{(m)}(t) = - \frac{i \phi_\mu^*(0)}{2\pi} \int_0^{2\pi} d\theta \, e^{-i m\theta}
\frac{\mathcal{E}_\mu^{(1)} + \mathcal{E}_\mu^{(2)} e^{i\theta}}{\left|\Omega_\mu^{(1)} + \Omega_\mu^{(2)} e^{i\theta}\right|}
\sin\left(\left|\Omega_\mu^{(1)} + \Omega_\mu^{(2)} e^{i\theta}\right| t\right),
\end{equation}
where $\mathcal{E}^{(i)} = d_\mu E_\mu^{(i)}$ and the frequencies
$\Omega_\mu^{(1,2)}$ are defined as the amplitudes of
$\Omega_\mu(\mathbf{x})$ corresponding to different pulses,
$\Omega_\mu^{(i)} = 2\sqrt{\sum_\sigma |\mathcal{E}^{(i)}_{\sigma,
s_\mu}|^2}$. 

\begin{figure}
  \includegraphics[width=6in]{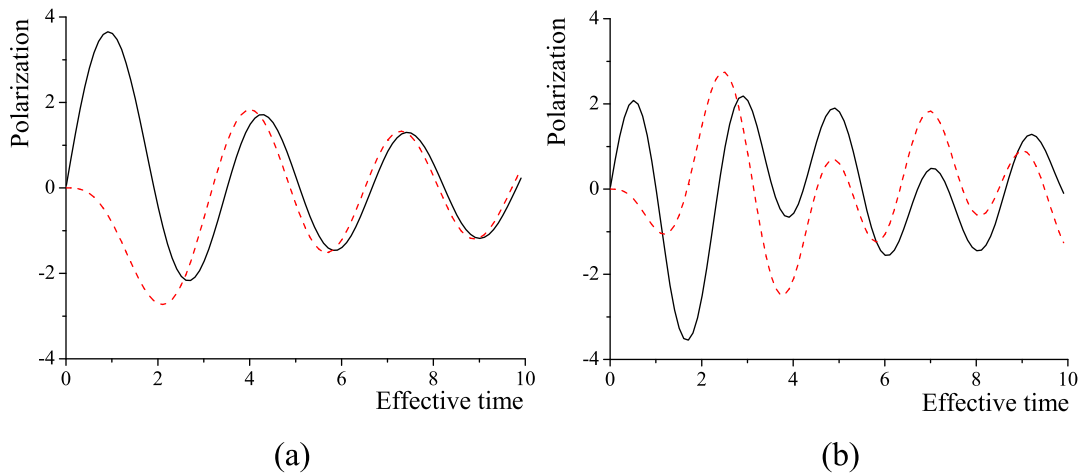}
  \caption{(Color online) The exciton polarization in the case
  when the excitation consists of two plane waves with
  non-collinear wave vectors.
  The solid and dashed lines show the response in the forward ($m=1$)
  and the four-wave mixing ($m=2$) directions for
  (a) $\Omega_\mu^{(1)}/\Omega_\mu^{(2)}=1$, (b) $\Omega_\mu^{(1)}/\Omega_\mu^{(2)}=2$.
 The effective time is defined as $\widetilde{t} = \Omega_\mu^{(2)} t$.
  }\label{fig:two_pulse_Rabi}
\end{figure}

In Fig.~\ref{fig:two_pulse_Rabi} we show the numerical evaluations of
Eq.~(\ref{eq:n_amplitudes_mwm}) for different ratios between the
amplitudes of the plane waves constituting the excitation along one of the
forward and four-wave mixing directions. The initial stage of the dynamics
for $t\to0$ can be obtained directly from series~(\ref{eq:P_mu_sum_A}) in
the form (for $m > 0$)
\begin{equation}\label{eq:pkn_small_t}
 P_{\mu}^{(m)}(t) = i \frac{(-1)^{m} \phi_\mu^*(0)}{(2m - 1)!} \,
\frac{\mathcal{E}_\mu^{(2)}}{ \eta_\mu}
 \left(\eta_\mu t\right)^{2m-1},
\end{equation}
where $\eta_\mu = \sqrt{\Omega_\mu^{(1)} \Omega_\mu^{(2)}}$.

The evolution of the exciton polarization in the opposite limit $t\to
\infty$ is found from integral representation~(\ref{eq:n_amplitudes_mwm})
using the stationary phase method
\begin{equation}\label{eq:pkn_big_t}
\begin{split}
 P^{(m)}_\mu(t) = -i \frac{\phi_\mu^*(0)}{\sqrt{2\pi \eta_\mu^2 t}}
 &\left\{ \mathrm{sign}(\mathcal{E}_\mu^{(1)} - \mathcal{E}_\mu^{(2)})(-1)^m
 \frac{|\mathcal{E}_\mu^{(1)} - \mathcal{E}_\mu^{(2)}|}{\sqrt{\left|\Omega_\mu^{(1)} - \Omega_\mu^{(2)}\right|}}
 \sin\left[\left|\Omega_\mu^{(1)} - \Omega_\mu^{(2)}\right|t +
 \frac{\pi}{4}\right]
 \right.\\
 & \left. +
 \frac{\mathcal{E}_\mu^{(1)} + \mathcal{E}_\mu^{(2)}}{\sqrt{\Omega_\mu^{(1)} + \Omega_\mu^{(2)}}}
 \sin\left[\left(\Omega_\mu^{(1)} + \Omega_\mu^{(2)}\right)t -
 \frac{\pi}{4}\right]
 \right\}.
\end{split}
\end{equation}
It shows that in this limit the $P^{(m)}_\mu(t)$ demonstrates power decay
$\propto t^{-1/2}$ and is essentially a superposition of two harmonics
with frequencies $\left|\Omega_\mu^{(1)} \pm \Omega_\mu^{(2)}\right|$. Of
course, the limit $t\to\infty$ should not be taken too literally, since in
this limit the main assumption $\omega_\mu t \ll 1$ is no longer valid and
the effect of desynchronization of the MWM polarization of high orders
(see below) becomes important. However, if $\Omega_\mu^{(1)}$ and
$\Omega_\mu^{(2)}$ are not too close then after just a few oscillations
the dependence $P^{(m)}_\mu(t)$ is satisfactory approximated by its
asymptotic form. For example, the complex behavior of the exciton
polarization shown in Fig.~\ref{fig:two_pulse_Rabi}b is a result of
existence of two harmonics with the frequencies $\left|\Omega_\mu^{(1)}
\pm \Omega_\mu^{(2)}\right|$, which exist if $\Omega_\mu^{(1)} \ne
\Omega_\mu^{(2)}$.

If $\Omega_\mu^{(1)} = \Omega_\mu^{(2)} = \Omega_\mu$ then the slow
harmonic vanishes. Indeed, in this case, as follows from
Eq.~(\ref{eq:n_amplitudes_mwm}), MWM polarizations can be presented in the
form
\begin{equation}\label{eq:pkn_exact}
\begin{split}
  P^{(m)}_\mu(t) = i (-1)^m & \phi_\mu^*(0) \left[
 \frac{\mathcal{E}_\mu^{(2)}}{\Omega_\mu} J_{2m -1}(2 \Omega_\mu t) \right. \\
 - & \left.\left(\mathcal{E}_\mu^{(1)} - \mathcal{E}_\mu^{(2)}\right)
 \int_0^t dt' \, J_{2m}(2 \Omega_\mu t')
  \right],
\end{split}
\end{equation}
where $J_m(t)$ are the Bessel functions of the first kind. This solution
asymptotically oscillates with the frequency $2 \Omega_\mu$, which is
consistent with the simple picture of the Rabi frequency renormalized due
to doubling the external field.

The interesting difference between the Rabi oscillations in the case of a
single pulse excitation and the oscillations of the MWM polarization is
that the latter asymptotically decays in time $\propto t^{-1/2}$. It can
be understood taking into account the dynamical origin of formation the
MWM polarization. As follows from Eq.~(\ref{eq:n_amplitudes_mwm}) the MWM
polarizations $P_\mu^{(m)}(t)$ satisfy the equation of motion of a
classical tight-binding model
\begin{equation}\label{eq:TBM_polarizations}
  -\ddot{P}_\mu^{(m)}(t) = \widetilde{\Omega}^2_\mu P_\mu^{(m)}(t)
  + \eta_\mu^2 {P}_\mu^{(m+1)}(t) + \eta_\mu^2 P_\mu^{(m-1)}(t),
\end{equation}
where $\widetilde{\Omega}^2_\mu = {\Omega_\mu^{(1)}}^2 +
{\Omega_\mu^{(2)}}^2$. Equations~(\ref{eq:TBM_polarizations}) are
supplemented by the initial conditions $P_\mu^{(m)}(0) \equiv 0$ for all
$m$ and $\dot{P}_\mu^{(m)}(0) = 0$ for all $m$ except $m=0,1$, for which
one has $\dot{P}_\mu^{(0)}(0) = -i \mathcal{E}^{(1)} \phi_\mu^*(0)$ and
$\dot{P}_\mu^{(1)}(0) = -i \mathcal{E}^{(2)} \phi_\mu^*(0)$. This
representation allows one to estimate the effect of the dependence of the
exciton energy on the in-plane momentum. It leads to modification of the
``eigenfrequency", $\widetilde{\Omega}^2_\mu$, by the term $\sim (m\Delta
{K})^2/2m_\mu$. Even if $(\Delta {K})^2/2m_\mu \ll \widetilde{\Omega}_\mu$
this contribution becomes essential for MWM exciton polarizations of
significantly high orders. However, for the most important case of MWM
polarizations of low orders on the time scale of few Rabi flops the effect
of the correction due to the exciton dispersion can be neglected.

It follows from the representation~(\ref{eq:TBM_polarizations}) that
initially the excitation is localized at the sites $m=0,1$ (the directions
of linear response). With time the energy spreads along the chain thanks
to the coupling $\propto \eta_\mu^2$ between different sites. In terms of
the MWM polarization the dynamics of this spreading corresponds to
excitation MWM polarizations of higher and higher orders. The dynamical
model described by Eq.~(\ref{eq:TBM_polarizations}) has the first integral
\begin{equation}\label{eq:TBM_energy}
  \sum_{m} \left\lbrace \left| \dot{P}_\mu^{(m)}(t)\right|^2
 + \widetilde{\Omega}^2_\mu \left|{P}_\mu^{(m)}(t)\right|^2
 + \eta_\mu^2 {P_\mu^{(m)}}^*(t)
 \left[{P}_\mu^{(m-1)}(t) + {P}_\mu^{(m+1)}(t)\right] \right\rbrace
 = |\phi_\mu(0)|^2 \left(\left|\mathcal{E}_\mu^{(1)}\right|^2 +
 \left|\mathcal{E}_\mu^{(2)}\right|^2\right).
\end{equation}
From this perspective it is clear that the decay of the Rabi oscillations
is the consequence of spreading the excitation among MWM polarizations of
different orders.

\section{The effect of internal dynamics and the Coulomb interaction}
\label{sec:internal}

In the previous section we have studied in details the dynamics of the
exciton polarization in the short time limit, when the main contribution
into dynamics comes from the interaction with the external field. Here we
consider the effect of the internal semiconductor dynamics and study how
the results of the previous section appear in the framework of a more
general description. We approach this problem describing the dynamics of
the interband polarizations $p_{\sigma,s}(\mathbf{x}_1,\mathbf{x}_2) =
\aver{c_{s}(\mathbf{x}_2)v_\sigma(\mathbf{x}_1)}$, the electron-electron
$e_{s_1,s_2}(\mathbf{x}_1,\mathbf{x}_2) =
\aver{c^\dagger_{s_1}(\mathbf{x}_1)c_{s_2}(\mathbf{x}_2)}$ and the
hole-hole $h_{\sigma_1,\sigma_2}(\mathbf{x}_1,\mathbf{x}_2) =
\aver{v^\dagger_{\sigma_1}(\mathbf{x}_1)v_{\sigma_2}(\mathbf{x}_2)}$
correlation functions by the semiconductor Bloch equation (see
Appendix~\ref{app:SBE_coordinate_derivation}).

First we notice that the most significant contribution of the full
dynamics would be the production of the response, which was absent in the
approximation used in the previous Section. This would be generating the
interband polarization corresponding to dark states (class $C_3$). It can
be proven, however, that the dynamics described by
Eqs.~\eqref{eq:SBE_coordinate} does not support excitation of the dark
states. Indeed, integrating both sides of the dynamical equations with
respect to time in the interval $(0,\Delta t)$ with $\Delta t \to 0$ one
can see that initially the spin state of the interband polarization
belongs to $\mathcal{V}_{\mathcal{E}}$. Next, considering the integral
over the interval $(t, t+\Delta t)$ one can see that if at instant $t$ the
state of the system, i.e. $p$, $h$ and $e$, belongs to
$\mathcal{V}_{\mathcal{E}}$, $\mathcal{V}_h$ and $\mathcal{V}_e$,
respectively, then so does the state at $t +\Delta t$. It should be noted,
thereby, that in quantum wells the valence band mixing does not lead to a
violation of invariance of the spin classes since the Hamiltonian
describing the mixing is an element of $\mathcal{V}_h$. Thus, if initially
the system is in the ground state then the interband polarization
corresponding only to the bright states will be produced at least as long
as the SBE and the approximation of angular independence of the dipole
moment hold.

As the next step we consider the limit of very short time response such
that following the analysis of simple model~\eqref{eq:field-coupling} the
contribution of the kinetic energy terms can be neglected, while other
terms including non-linear are kept. The solution found in the previous
section, Eq.~\eqref{eq:p_mu}, suggests the interband polarization in the
form $p(\mathbf{x},\mathbf{y}) \propto \delta(\mathbf{x} - \mathbf{y})$.
Substituting this ansatz into Eqs.~\eqref{eq:SBE_coordinate} with
neglected kinetic energy terms and introducing a cut-off of the Coulomb
potential such that $V(0) = V_0 < \infty$ one arrives at the system of
equations independent of $V_0$, which is structurally similar to
Eqs.~\eqref{eq:amplitudes_kernels_dynamics} and which yields the same
result for the exciton polarization as Eq.~\eqref{eq:p_mu}. The
cancelation of the terms proportional to the Coulomb potential is
particularly evident from the representation of the SBE in
form~\eqref{eq:SBE_FULL}.

This consideration shows that the most significant correction to the
approximate solution results from the kinetic energy terms rather than
from the Coulomb interaction as one might expect considering that the
Coulomb interaction is responsible for the non-linear terms in the SBE.

We would like to emphasize at this point that the quantity of our main
interest is the exciton polarization $P_\mu$ since these are the excitons
that constitute true single-particle semiconductor states. As follows
Eq.~\eqref{eq:exciton_operator_elementary}, $P_\mu$ is related to the
interband polarization $p_{\sigma,s}$ through
\begin{equation}\label{eq:interband_exciton_relation}
  P_\mu = \int d\mathbf{x}d\mathbf{y}\, \phi^*_\mu(\mathbf{x}, \mathbf{y})
  p_{\sigma_\mu,s_\mu}(\mathbf{y}, \mathbf{x}).
\end{equation}
For bound exciton states this naturally introduces a characteristic
spatial scale, the exciton Bohr radius $r_B$. This circumstance together
with the assumption that the typical spatial variation of the external
excitation in the plane of the quantum well is small comparing to the Bohr
radius $K_{1,2}r_B \ll 1$ allows one to simplify the SBE using the
parametric approximation. For this we introduce new spatial variable
$\mathbf{R}=(\mathbf{x} + \mathbf{y})/2$ and $\mathbf{r} = \mathbf{x} -
\mathbf{y}$. Next we notice that in the case of single pulse or two pulse
excitations the solutions of the SBE are invariant with respect to either
infinitesimal or finite (by $\Delta \mathbf{K}/(2\pi \Delta K^2)$)
translations in $\mathbf{R}$ plane. Having in mind the consecutive
convolution of the solutions with the exciton wave function we can neglect
the terms $\propto \Delta K r_B$ and leave only the dependence on
$\mathbf{R}$ as a parameter.\cite{BINDER:1995,BINDER:1997,Haug_Koch} Thus,
assuming for simplicity that $\mathbf{K}_1=0$ we approximate
Eqs.~\eqref{eq:SBE_coordinate} by
\begin{equation}\label{eq:SBE_parametric}
  \begin{split}
  i\dot{p}_{\sigma,s}(\mathbf{r}, \mathbf{R}) & = \widetilde{K}_{\sigma,s}[p] +
\mathcal{E}_{\sigma,s}(\mathbf{R})\delta(\mathbf{r})
  - \int d\mathbf{r}' \left[
  \widetilde{\mathcal{E}}_{\sigma',s}(\mathbf{r}',\mathbf{R})h_{\sigma',\sigma}(\mathbf{r}' - \mathbf{r},\mathbf{R})
  - \widetilde{\mathcal{E}}_{\sigma,s'}(\mathbf{r}',\mathbf{R})e_{s',s}(\mathbf{r}' + \mathbf{r},\mathbf{R})
  \right] , \\
 -i \dot{e}_{s_1,s_2}(\mathbf{r}, \mathbf{R}) = & \,
 \widehat{K}_{s_1,s_2}[e] + \int d\mathbf{r}' \left[
  \widetilde{\mathcal{E}}^*_{\sigma',s_1}(\mathbf{r}',\mathbf{R})p_{\sigma',s_2}(\mathbf{r}' + \mathbf{r},\mathbf{R})
  -
  \widetilde{\mathcal{E}}_{\sigma',s_2}(\mathbf{r}',\mathbf{R})p^*_{\sigma',s_1}(\mathbf{r}' - \mathbf{r},\mathbf{R})
  \right], \\
 -i \dot{h}_{\sigma_1,\sigma_2}(\mathbf{r}, \mathbf{R}) = &\,
 \widehat{K}_{\sigma_1,\sigma_2}[h] + \int d\mathbf{r}' \left[
  \widetilde{\mathcal{E}}^*_{\sigma_1,s'}(\mathbf{r}',\mathbf{R})p_{\sigma_2,s'}(-\mathbf{r} - \mathbf{r}',\mathbf{R})
  -
  \widetilde{\mathcal{E}}_{\sigma_2,s'}(\mathbf{r}',\mathbf{R})p^*_{\sigma_1,s'}(\mathbf{r} + \mathbf{r}',\mathbf{R})
    \right].
  \end{split}
\end{equation}
Here $\widetilde{\mathcal{E}}^*_{\sigma,s}(\mathbf{r},\mathbf{R}) =
\mathcal{E}_{\sigma,s}(\mathbf{R})\delta(\mathbf{r}) -
V(\mathbf{r})p_{\sigma,s}(\mathbf{r},\mathbf{R})$ and the
integro-differential operators in Eqs.~\eqref{eq:dynamical_operators} are
substituted by
\begin{equation}\label{eq:integro_diff_parametric}
   \begin{split}
  \widehat{H}_{s_1,s'}f_{s'}=  & \left[-\frac{1}{2m_{s_1}} \frac{\partial^2}{\partial \mathbf{r}^2}
   + H_{s_1}\right]f_{s_1}(\mathbf{r},\mathbf{R})
    -\int d\mathbf{r}'\, V(\mathbf{r}')e_{s_1,s'}(\mathbf{r}',\mathbf{R})
    f_{s'}(\mathbf{r} - \mathbf{r}',\mathbf{R}), \\
  \widehat{H}_{\sigma_1,\sigma'}f_{\sigma'}= &
  -\frac{1}{2m_{\sigma_1,\sigma'}} \frac{\partial^2}{\partial \mathbf{r}^2} f_{\sigma'}(\mathbf{r}, \mathbf{R})
    + H_{\sigma_1}f_{\sigma_1}(\mathbf{r},\mathbf{R})
    -\int d\mathbf{r}'\, V(\mathbf{r}')h_{\sigma_1,\sigma'}(\mathbf{r}',\mathbf{R})
    f_{\sigma'}(\mathbf{r} - \mathbf{r}',\mathbf{R}).
  \end{split}
\end{equation}
System~\eqref{eq:SBE_parametric} in the short time limit when all terms in
the right-hand side but those depending on the external field are
neglected reproduces the results obtained in the previous section from
Eq.~\eqref{eq:P_t_series_presentation}. In particular the multi-wave
mixing response is obtained by the Fourier transform over the parameter
$\mathbf{R}$ similarly to Eq.~\eqref{eq:n_amplitudes_mwm}. In what follows
we will use Eqs.~\eqref{eq:SBE_parametric} in even simpler form assuming
the single-pulse excitation only and thus omitting the dependence on
$\mathbf{R}$. It suffices for our purpose of discussion of the effect of
the internal dynamics and the Coulomb interaction. We also omit the
off-diagonal elements of the holes Hamiltonian
$\widehat{H}_{\sigma_1,\sigma_2}$ describing the mixing of the valence
bands. Thus we characterize these bands by $m_{hh}$ and $m_{lh}$ that is
by the masses of the heavy- and light-holes, respectively.

Considering the terms besides the kinetic energy in the right-hand side of
these equations as sources the responses are governed by the free-particle
propagators for the correlation functions and the Coulomb propagator for
the interband polarization. The fundamental property of these propagators
is that in the short time limit $t\to 0$ they turn to spatial
$\delta$-functions owing to fast spatial oscillations of the kernel
\begin{equation}\label{eq:free_propagator}
  K(\mathbf{r}_1, \mathbf{r}_2; t) = \sqrt{\frac{m}{2\pi i t}}
  \exp\left[i m |\mathbf{r}_1 - \mathbf{r}_2|^2/2t\right].
\end{equation}
In order to evaluate the parameter $m$ we take into account that the
evolution of the semiconductor initially being in the ground state is
spanned by the spin classes $C_{1,2}$. Thus for the electron correlation
function $m = \infty$ so that the propagator remains local. For the hole
correlation functions this parameter is finite for the off-diagonal
elements $m = m_\Delta = (m_{lh}^{-1} - m_{hh}^{-1})^{-1}$. For the
interband polarization in the short time limit the Coulomb propagator can
be approximated by the free-particle
propagator\cite{BLINDER:1984,KUNIKEEV:2000} with $m \approx
m_{xh}=(m_e^{-1}+m_{hh})^{-1}$ and $m \approx
m_{xl}=(m_e^{-1}+m_{lh})^{-1}$, where $m_{x\sigma}$ are the excitons'
reduced masses and $m_e$ is the electron mass. The effect of the
dispersion of the electrons and holes on the \emph{exciton} polarization
is expressed as a distortion of the exciton wave function under the action
of the free-particle propagators. The time scale, at which the distortion
becomes essential, can be estimated as the time required for the initial
$\delta$-shape of the kernel to acquire the width of the order of the
exciton Bohr radius. This leads to the estimate
\begin{equation}\label{eq:free_dispersion_time}
  t_c \lesssim \frac {m r_B^2} {4 \pi}.
\end{equation}
The numerical value of the typical time scale for GaAs is determined by
$m_{hh}=0.45 m_0$, $m_{lh} = 0.082 m_0$, $m=m_\Delta$ and $r_B = r_{xh} =
\epsilon_b /2 m_{xh} e^2$, where $m_0$ is the electron mass in empty space
and $\epsilon_b = 13$ is the background dielectric function. Using these
values in Eq.~\eqref{eq:free_dispersion_time} we find $t_c \sim 20$ fs.
Thus for pulses with duration shorter than $20$ fs and the intensity
sufficiently high to produce several Rabi flops, the approximation used in
the previous Section is satisfactory. However, as will be shown
below the exciton polarizations corresponding to the states in the
discrete spectrum are less sensitive to the deviation of the dynamics from
its short time limit.

First we discuss the general effect of the electron-hole dispersion on the
Rabi oscillations. For this we consider the case of excitation by a single
circularly polarized pulse of high intensity. For such excitation we
neglect the contribution of the non-linear Coulomb terms (see below) and
additionally simplify the internal dynamics assuming the effects of the
detuning and the energy offsets to be small at the time scales under
consideration. Since in this case the translational invariance in the
plane of the quantum well is restored it is more convenient to solve
Eqs.~\eqref{eq:SBE_parametric} rewriting them in the momentum
representation.\cite{BINDER:1995,BINDER:1997,Haug_Koch} The Laplace
transformation $p(\lambda) = \int_0^\infty p(t) e^{-\lambda t}dt$ of the
solution is found to be
\begin{equation}\label{eq:interband_dispersion_frequency}
  p_{\sigma, s}(\lambda, \mathbf{k}) = -i \frac{\mathcal{E}_{\sigma,s}}{(2\pi)^2}
  \cdot \frac {\lambda - i\omega_{\sigma,s}(\mathbf{k})}{
  \lambda[\lambda^2 + \omega^2_{\sigma,s}(\mathbf{k}) + \Omega^2_{\sigma,s}]}
\end{equation}
where $\omega_{\sigma,s}(\mathbf{k}) = k^2/2 m_{x\sigma}$ and
$\Omega^2_{\sigma,s} = 4 (|\mathcal{E}_{\sigma,s}|^2 +
|\mathcal{E}_{\bar{\sigma},s}|^2)$ with $\bar{\sigma}$ being the hole spin
state complimentary to $\sigma$ in the respective spin class. In the case
under consideration because of circular polarization of the excitation
pulse only interband polarization corresponding to the same helicity is
created, thus, if $\mathcal{E}_{\sigma,s} \ne 0$ then
$\mathcal{E}_{\bar{\sigma},s} = 0$ and vice versa. We, however, write down
the ``full" expression for $\Omega_{\sigma,s}$ in order to show the
relation with Eq.~\eqref{eq:RabiF_x}. The poles of the right-hand side of
Eq.~\eqref{eq:interband_dispersion_frequency} determine the characteristic
frequencies yielding the time dependence in the form
\begin{equation}\label{eq:interband_dispersion_time}
 \begin{split}
   p_{\sigma, s}(t, \mathbf{k}) & = - i
   \frac{\mathcal{E}_{\sigma,s}}{(2\pi)^2 \widetilde{\Omega}_{\sigma,s}(\mathbf{k})}
 \left\{ \sin[\widetilde{\Omega}_{\sigma,s}(\mathbf{k}) t] -
 2i \frac {\omega_{\sigma,s}(\mathbf{k})} {\widetilde{\Omega}_{\sigma,s}(\mathbf{k})}
 \sin^2 [\widetilde{\Omega}_{\sigma,s}(\mathbf{k}) t/2])\right\}, \\
 e_{s,s}(t, \mathbf{k}) & =
 h_{\sigma,\sigma}(t, \mathbf{k}) =
 \frac{4|\mathcal{E}|^2_{\sigma,s}}{(2\pi)^2\widetilde{\Omega}^2_{\sigma,s}(\mathbf{k})}
 \sin^2 [\widetilde{\Omega}_{\sigma,s}(\mathbf{k}) t/2]
\end{split}
\end{equation}
where $\widetilde{\Omega}_{\sigma,s}(\mathbf{k}) =
\sqrt{\Omega^2_{\sigma,s} + \omega_{\sigma,s}(\mathbf{k})^2}$. It is
interesting to note the similarity to the simple dynamical models
discussed in Introduction [see Eq.~\eqref{eq:field_coupling_solution}].

\begin{figure}
  \includegraphics[width=6in]{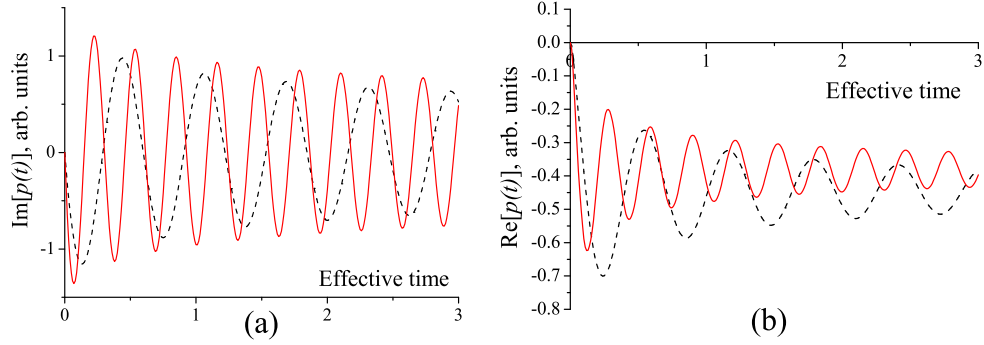}\\
  \caption{The effect of the electron-hole dispersion on the
  Rabi oscillations. The time dependence of the real part of the exciton
  polarization is shown for $\Omega_{\sigma,s}/\omega_{\sigma,s}(1/r_B) = 5$ (solid line)
  and $\Omega_{\sigma,s}/\omega_{\sigma,s}(1/r_B) = 10$ (dashed line).
}\label{fig:exciton_dispersion}
\end{figure}

The \emph{exciton} polarization is found by convoluting
Eq.~\eqref{eq:interband_dispersion_time} with the exciton wave function
$\phi_{\sigma,s}(\mathbf{k}) = 2 \sqrt{2\pi} r_{\sigma,s}(1 + k^2
r_{\sigma,s}^2)^{-3/2}$ according to
Eq.~\eqref{eq:interband_exciton_relation}. In
Fig.~\ref{fig:exciton_dispersion} we plot the time dependence of the
exciton polarization for different intensities of the external excitation.
Since the exciton wave function mostly localized inside the region $k \sim
1/r_B$ and by assumption $\omega_{\sigma,s}(1/r_B) < \Omega_{\sigma,s}$
this convolution does not lead to appearance of new frequencies, thus the
oscillations of the exciton polarization are characterized by a slightly
modified Rabi frequency $\Omega_{\sigma,s}$.

\begin{figure}
  \includegraphics[width=3in]{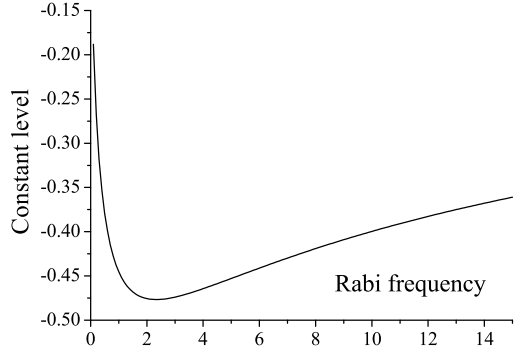}\\
  \caption{The dependence of the constant level on the
  Rabi frequency $\Omega_{\sigma,s}/\omega_{\sigma,s}(1/r_B)$.
  }\label{fig:level_absolute}
\end{figure}

There are two significant corrections to the picture of the oscillations
obtained in the previous section. First, there is slower than exponential
decay due to dephasing of the components corresponding to different $k$.
The decay can be estimated considering the long time asymptotic
$t\to\infty$. Since the phase in Eq.~\eqref{eq:interband_dispersion_time}
reaches its stationary points at the ends of the $k$ interval, i.e. at $k=0$, 
it implies the asymptotic power decay of the amplitude as $1/t$
independent on the intensity of the excitation. The second modification is
related to the second term in the brackets in
Eq.~\eqref{eq:interband_dispersion_time}
[Fig.~\ref{fig:exciton_dispersion}b]. This term has the form of
oscillations with the Rabi frequency $\Omega_{\sigma,s}$ near a constant
level, which position strongly depends on the intensity of the excitation
for $\Omega_{\sigma,s} \lesssim 2 \omega_{\sigma,s}(1/r_B)$ and slowly
decays towards $0$ for higher values of the Rabi frequency as shown in
Fig.~\ref{fig:level_absolute}.

Solutions~\eqref{eq:interband_dispersion_time} allow us to estimate the
effect of the Coulomb interaction and to show that the contribution of the
non-linear terms into the right-hand side of equations of
motion~\eqref{eq:SBE_parametric} vanishes with increasing the intensity of
the excitation. Physically this is supported by, first, observation that
the interband polarization and the single particle correlation functions
as functions of the external excitation are limited from above and, hence,
the nonlinear terms are scaled as the exciton binding
energy\cite{OSTREICH:1993}, while the other terms increase with increasing
the intensity of the external field. Second, one can also apply the result
obtained from Eq.~\eqref{eq:SBE_FULL} about vanishing the non-linear terms
in the limit of $\delta$-functional profile of the interband polarization
and the single-particle correlation functions. As follows from
Eqs.~\eqref{eq:interband_dispersion_time} with increasing the intensity of
the excitation the spatial distribution of $p_{\sigma,s}$,
$h_{\sigma,\sigma}$ and $e_{s,s}$ narrows with the characteristic width
$\propto 1/k_\Omega$, where $k_\Omega =
\sqrt{2m_{x\sigma}\Omega_{\sigma,s}}$.

\begin{figure}
  \includegraphics[width=3in]{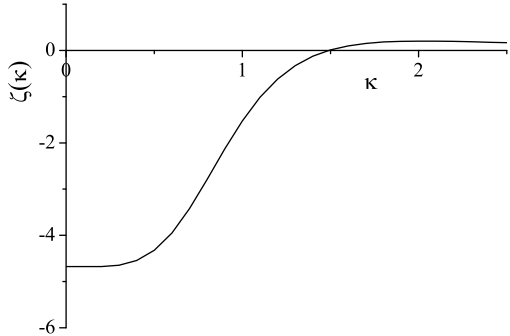}\\
  \caption{The dependence on $\kappa = k/k_\Omega$ of the integral in
  Eq.~\eqref{eq:zeta_def} describing the effective
  modification of the source due to the nonlinear terms
  in the limit of high intensity of the external field.}\label{fig:integral_kappa}
\end{figure}

There are different ways to estimate the contribution of the non-linear
Coulomb terms into the dynamics. In order to illustrate the application of
the physical ideas mentioned above we consider the contribution into the
right-hand side of the equation of motion for the interband polarization
coming from the term written in the second line in
Eq.~\eqref{eq:SBE_FULL}. In the simplest case of a single pulse excitation
this term yields an addition to the equation with respect to
$p_{\sigma,s}(t,\mathbf{k})$ in the form
\begin{equation}\label{eq:nonlinear_addition_pk}
   \Delta S_{\sigma,s}(\mathbf{k}) =  \int d\mathbf{k}' V(\mathbf{k} - \mathbf{k}')
   \left[p_{\sigma,s}(\mathbf{k}') n(\mathbf{k})
   - p_{\sigma,s}(\mathbf{k}) n(\mathbf{k}') \right],
\end{equation}
where $n(\mathbf{x}) = e_{s,s}(\mathbf{x}) +
h_{\sigma,\sigma}(\mathbf{x})$ and we have taken into account that the
dynamics is restricted to a particular spin class. We illustrate the
typical line of calculations leaving only the first term in the expression
for $p_{\sigma,s}(\mathbf{k})$ and approximating the oscillating terms by
$1$ thus presenting
\begin{equation}\label{eq:nonlinear_addition_pk1}
   \Delta S_{\sigma,s}(\mathbf{k}) \sim -i \frac{8\mathcal{E}^3_{\sigma,s}}{(2\pi)^4}
   \int d\mathbf{k}' \frac{V(\mathbf{k} - \mathbf{k}')}{\widetilde{\Omega}^2_{\sigma,s}(\mathbf{k})
   \widetilde{\Omega}^2_{\sigma,s}(\mathbf{k}')}
   \left[\widetilde{\Omega}_{\sigma,s}(\mathbf{k}) -
   \widetilde{\Omega}_{\sigma,s}(\mathbf{k}')\right].
\end{equation}
Rescaling the wave vectors and using for the single-pulse excitation
$\Omega_{\sigma,s} = 2\mathcal{E}_{\sigma,s}$ we obtain
\begin{equation}\label{eq:nonlinear_addition_scale}
  \Delta S_{\sigma,s}(\mathbf{k}) \sim -i \frac{1}{(2\pi)^4}
  \sqrt{\frac{\epsilon_{x\sigma}}{\Omega_{\sigma,s}}} \zeta\left(\frac{\mathbf{k}}{k_\Omega}\right)
\end{equation}
where $\epsilon_{x\sigma}= 2m_{x\sigma}e^4/\epsilon_b^2$ is the exciton
binding energy and
\begin{equation}\label{eq:zeta_def}
  \zeta(\boldsymbol{\kappa}) =
  \int d \boldsymbol{\kappa}' \frac{\sqrt{1+ \kappa^4} - \sqrt{1+ \kappa'^4}}{(1+ \kappa^4)(1+ \kappa'^4)
  |\boldsymbol{\kappa} - \boldsymbol{\kappa}'|}.
\end{equation}
The dependence of this function on $\kappa$ is shown in
Fig.~\ref{fig:integral_kappa}. It varies slowly over the interval $\kappa
< 1/2$, where its magnitude is close to $\zeta(0) =
\pi^2/\sqrt{2} - 8\Gamma^2(5/4)\sqrt{\pi}\approx -4.67$. Thus, if the Rabi
frequency exceeds the exciton binding energy or, equivalently, if the
typical spatial scale of the interband polarization $1/k_\Omega$ is
smaller than the exciton Bohr radius the effect of the non-linear terms on
the time dependence of the exciton polarization reduces to a relatively
small modification of the amplitude of the external field.

This argument allows one to reformulate the criterion of validity of the
short time approximation for the exciton polarization in terms of relation
between the exciton Bohr radius $r_B$ and the typical spatial width of the
\textit{interband} polarization due to the interaction with the external
field $\sim 1/k_\Omega$. If $r_B k_\Omega \gg 1$ then the result of
convolution of the exciton wave function with the interband polarization
$p_{\sigma,s}$ does not depend on the details of the dependence of
$p_{\sigma,s}(\mathbf{k})$ on $\mathbf{k}$. Indeed, since in this case the
main contribution into the integral $P=\int
d\mathbf{x}\,\psi^*(\mathbf{x}) p(\mathbf{x})$ comes from the small area
near $\mathbf{x} = 0$ we can approximate the integral by
\begin{equation}\label{eq:P_approximation}
  P \approx \psi^*(0) \int d\mathbf{x} p(\mathbf{x}) = (2\pi)^2 \psi^*(0)
  p(\mathbf{k}=0)
\end{equation}
which turns Eq.~\eqref{eq:interband_dispersion_time} into
Eq.~\eqref{eq:plane_wave_excitation_h} exactly (including the cancelation
of the factor $(2\pi)^2$). This consideration shows that the exciton
polarization corresponding to the states in the discrete spectrum is less
sensitive to the deviations from the short-time dynamics of the system
than the interband polarization. The latter contains information about all
exciton states including those from the continuum spectrum whose spatial
variation is no longer characterized by the exciton Bohr radius. The
condition $r_B k_\Omega = \sqrt(\Omega_{\sigma,s}/\epsilon_{x\sigma}) \gg
1$ imposes the limitation on the intensity of the external field that the
period of the Rabi oscillations should be smaller than $\sim 200$ fs (for
GaAs). 

An interesting effect of the dispersion is revealed when the excitation
pulse is not circularly polarized. In this case both states constituting a
spin class are involved in the dynamics resulting in the appearance of a
new characteristic frequency. In order to illustrate this effect we
consider the case when the intensity of the excitation is significantly
high so that the effect dispersion of the hole-hole correlation function,
which is quantified by the parameter $m_\Delta^{-1}=m_{lh}^{-1} -
m_{hh}^{-1}$, can be treated perturbatively. In order to simplify the
expression we assume additionally that both heavy- and light-hole excitons
are described by the same dispersion law $\omega_{\sigma,s}(\mathbf{k})$.
Under these assumptions the Laplace transform of the interband
polarization can be written as
\begin{equation}\label{eq:two-frequencies}
  p_{\sigma,s}(\lambda, \mathbf{k}) = p^{(0)}_{\sigma,s}(\lambda, \mathbf{k}) +
  \frac{1}{2\pi^2 \lambda}
  \cdot \frac{\mathcal{E}_{\sigma,s} |\mathcal{E}_{\bar{\sigma},s}|^2 \omega_{\bar{\sigma},\sigma}(\mathbf{k})}{
  [\lambda^2 + \widetilde{\Omega}^2_{\sigma,s}(\mathbf{k})]
  [(\lambda  + i \omega_{\sigma,s}(\mathbf{k})/2)^2 + \widetilde{\Omega}^2_{\sigma,s}(\mathbf{k})/4]},
\end{equation}
where $p^{(0)}_{\sigma,s}(\lambda, \mathbf{k})$ is given by
Eq.~\eqref{eq:interband_dispersion_frequency} and
$\omega_{\bar{\sigma},\sigma}(\mathbf{k}) = k^2/2 m_{\bar{\sigma},\sigma}$
with $m_{\bar{\sigma},\sigma}^{-1} = m_{\bar{\sigma}}^{-1} -
m_{\sigma}^{-1}$, i.e. $|m_{\bar{\sigma},\sigma}| = m_\Delta$. The
perturbation term in the right-hand side of Eq.~\eqref{eq:two-frequencies}
has poles at the same Rabi frequency
$\widetilde{\Omega}^2_{\sigma,s}(\mathbf{k})$ as in the case of the
circularly polarized excitation and at the frequency $\approx
\widetilde{\Omega}^2_{\sigma,s}(\mathbf{k})/4$ yielding a component of the
exciton polarization oscillating with doubled period. We would like to
emphasize that the new frequency was initially present among the
eigenfrequencies of the bright exciton classes [see
Eq.~\eqref{eq:c12_states}]. Its appearance, however, in the time evolution
of the exciton polarization was prohibited in the dispersionless limit by
the symmetry ${p^{(2)}}^* \leftrightarrow {p^{(1)}}$ of dynamical
equations~\eqref{eq:polarization_amplitudes}. The exciton finite masses
break the symmetry allowing for the frequency
${\Omega}^2_{\sigma,s}(\mathbf{k})/4$ to contribute to the evolution. With
this regard it could be speculated that if the dark excitons would be
involved in the dynamics their presence could be traced by admixture of
the frequencies specific for the dark exciton spin class $C_3$ [see
Eq.~\eqref{eq:c3_states}].

It should be emphasized that neither effect of decay of the Rabi
oscillations nor the admixture of ``extra" frequencies are present in
atomic analogues of the semiconductor Rabi oscillations. The decay is
essentially caused by the spatial spread of the true (single-particle)
eigenstates of the system. The appearance of additional frequencies is
related to the existence of different kinds of holes with continuous
spectra characterized by different masses.

%
%

\section{Conclusion}

We have studied the short time semiconductor response with respect to
resonant high intensity excitations. The main results are obtained
neglecting the slow dynamics, that is on the time scale provided by
detuning. We have calculated exactly the excitation polarization $P_\mu =
\bra{\psi(t)}B_\mu\ket{\psi(t)}$ in the coherent limit. We have found that
if the semiconductor is excited by a field consisting of one or two plane
waves the polarization of bright excitons demonstrates Rabi oscillations,
while the polarization of dark excitons identically zero. The main
difference between the single-wave and two-wave excitations is that in the
latter case the Rabi oscillations of the multi-wave mixing (MWM)
polarizations are generally a superposition of two harmonics ($\Omega_1 >
\Omega_2$) unless the amplitudes of the waves satisfy a special relation.
For example, if both waves are circularly polarized with the same helicity
then $\Omega_1 - \Omega_2 \sim 2dE$, where $E$ is amplitude of the weaker
wave, and the slow harmonic $\Omega_2$ collapses only if the amplitudes of
the both waves are the same. Also, in the case of the two pulse excitation
oscillations of the exciton polarization decay $\propto t^{-1/2}$ owing to
redistribution of the excitation over the multi-wave mixing polarizations.

In order to establish the dynamical origin of the vanishing polarization
of the dark states we considered the operator dynamics governed by the
external field. We found that there exist three invariant spin classes,
which do not mix with the evolution of the system. Two of these classes
correspond to bright excitons and the third one contains all the dark
states. It has turned out that the operator dynamics is described by six
frequencies. Only two of those frequencies show up as the Rabi frequencies
in the time evolution of the exciton polarizations (one for each bright
classes), while others are prohibited by a hidden symmetry.

The short time approximation used in the first part of the paper appears
naturally in the more general context of the semiconductor Bloch equation
(SBE). We use the SBE in order to discuss the effect of the Coulomb
interaction and the effect of the continuous spectra of the electrons and
holes on the results obtained in the limit of immediate response with
respect to intensive excitation. We found that if the system is initially
in the ground state then the SBE preserves the invariant spin classes.
Thus, it proves that as long as the SBE holds and the approximation of
angular independent dipole moment is justified the dark excitons do not
appear in the dynamics.

An interesting result of studying the short time approximation within the
context of the SBE is that the Coulomb interaction does not lead to
essential changes if the intensity of excitation is sufficiently high
compared to the exciton binding energy. Moreover, including the nonlinear
exchange terms directly to the equations of the short time limit (i.e.
without the kinetic energy) adds nothing new since they cancel each other.
In turn, the effects of the continuous spectrum of the electrons and holes
are much more significant. We show that the electron-hole dispersion while
does not change significantly the frequency of the oscillations of the
exciton polarization leads to the power law decay of the oscillations. An
interesting effect is related to the mass difference between heavy- and
light-holes. It breaks the symmetry that selects only two Rabi frequencies
out of six that are introduced into the system by the external field and
leads to admixture of the second nontrivial frequency for a specific
bright class.

\acknowledgments

We would like to thank Lu Sham and Cristiano Ciuti for useful discussions.
We acknowledge partial support through the NSF under Grant No. ECCS
0725514 and through the DARPA/MTO Young Faculty Award under Grant No.
HR0011-08-1-0059.

\appendix

\section{The Heisenberg representation induced by the external field}
\label{app:external_field_algebra}

In order to derive Eqs.~\eqref{eq:amplitudes_dynamics} one can either
consider a regularization of the $\delta$-function, which appears in the
anticommutation relations and in the right-hand-side of
Eq.~\eqref{eq:F_operator} or consider a more general form of
Eq.~\eqref{eq:F_operator} with the amplitudes being scalar functions,
which are convoluted with the electron and hole operators,
\begin{equation}\label{eq:general_F_operator}
   \widehat{F}(t) = A(t) + \sum_{\sigma_1,\sigma_2}\int d\mathbf{x} d\mathbf{y} \,
   g_{\sigma_1, \sigma_2}(\mathbf{x},\mathbf{y};t)
   v^\dagger_{\sigma_1}(\mathbf{x})v_{\sigma_2}(\mathbf{y}) + \ldots.
\end{equation}
Furthermore the Heisenberg equation of motion produces the dynamical
equations with respect to the amplitudes
\begin{equation}\label{eq:amplitudes_kernels_dynamics}
  \begin{split}
  \dot{f}_{s_1,s_2}(\mathbf{x},\mathbf{y}) = & \,
  i\mathcal{E}_{\sigma',s_1}(\mathbf{x})\pi^{(2)}_{\sigma',s_2}(\mathbf{x},\mathbf{y})-
                i\pi^{(1)}_{\sigma',s_1}(\mathbf{y},\mathbf{x})\mathcal{E}^*_{\sigma',s_2}(\mathbf{y}), \\
 \dot{g}_{\sigma_1,\sigma_2}(\mathbf{x},\mathbf{y})= &\,
            i \mathcal{E}_{\sigma_1,s'}(\mathbf{x})\pi^{(2)}_{\sigma_2,s'}(\mathbf{y},\mathbf{x})
        - i\pi^{(1)}_{\sigma_1,s'}(\mathbf{x},\mathbf{y})\mathcal{E}^*_{\sigma_2,s'}(\mathbf{y}), \\
  \dot{\pi}^{(1)}_{\sigma,s}(\mathbf{x},\mathbf{y}) = &\,
  -i \mathcal{E}_{\sigma,s'}(\mathbf{x}) f_{s,s'}(\mathbf{y},\mathbf{x}) -
            i g_{\sigma, \sigma'}(\mathbf{x},\mathbf{y}) \mathcal{E}_{\sigma',s}(\mathbf{y}), \\
  \dot{\pi}^{(2)}_{\sigma,s}(\mathbf{x},\mathbf{y}) = &\,
  i \mathcal{E}^*_{\sigma,s'}(\mathbf{x}) f_{s',s}(\mathbf{x},\mathbf{y}) +
            i g_{\sigma', \sigma}(\mathbf{y},\mathbf{x}) \mathcal{E}^*_{\sigma',s}(\mathbf{y}), \\
    \dot{A} = &\, i \int d\mathbf{x} \left[\pi^{(1)}_{\sigma',s'}(\mathbf{x},\mathbf{x})\mathcal{E}^*_{\sigma',s'}(\mathbf{x})
    - \pi^{(2)}_{\sigma',s'}(\mathbf{x},\mathbf{x})\mathcal{E}_{\sigma',s'}(\mathbf{x})\right],
  \end{split}
\end{equation}
where the summation over dashed spin variables is implied. The last
equation gives the time dependence of $\aver{\widehat{F}(t)}$ (compare
with Eq.~\eqref{eq:A_kernels}). Considering the dynamics of the diagonal,
$\mathbf{x}=\mathbf{y}$, values of the amplitudes in the case of spatially
homogeneous excitation $\mathcal{E}_{\sigma,s}(\mathbf{x}) =
\mathcal{E}_{\sigma,s}$ one obtains Eq.~\eqref{eq:amplitudes_dynamics}.

It should be emphasized that Eqs.~\eqref{eq:amplitudes_kernels_dynamics}
are not a semiconductor Bloch equations. These are equations with respect
to amplitudes of different operators entering the Heisenberg
representation of the operator $\widehat{F}(t)$. The semiconductor Bloch
equation, in turn, is written for average values of the specific operators
[see e.g. Eq.~\eqref{eq:SBE_coordinate}]. While these equations have
similar structure (in the linear approximation for the SBE) they are
different as can be see comparing closely
Eqs.~\eqref{eq:amplitudes_kernels_dynamics} and~\eqref{eq:SBE_coordinate}.
In order to find an average value of a particular operator using
Eqs.~\eqref{eq:amplitudes_kernels_dynamics} one needs to solve these
equations for specific initial conditions and then use this solution in
the last equation of Eqs.~\eqref{eq:amplitudes_kernels_dynamics}.

\section{The semiconductor Bloch equation in the coordinate representation}
\label{app:SBE_coordinate_derivation}

In the main text we consider the semiconductor response under the action
of an external field, which is spatially inhomogeneous in the plane of the
quantum well. Because the system does not posses the translational
invariance it is convenient to use the semiconductor Bloch equation in the
coordinate representation.

The Hamiltonian of an excited semiconductor in the rotating wave
approximation has the form
\begin{equation}\label{eq:semiconductor_Hamiltonian}
  H = H_{el} + H_h + H_C + H_e,
\end{equation}
where $H_{el}$ and $H_h$ are the standard electron and hole
single-particle Hamiltonians\cite{BINDER:1995,Haug_Koch}, $H_C$ is the
Hamiltonian of the Coulomb interaction and $H_e$ describes the
light-matter interaction
\begin{equation}\label{eq:light-matter_electron_hole}
  H_e = \int d\mathbf{x} \,\mathcal{E}_{\sigma,s}(\mathbf{x}) v^\dagger_\sigma(\mathbf{x})
  c^\dagger_s(\mathbf{x}) + \text{h.c.},
\end{equation}
where $\mathcal{E}_{\sigma,s}(\mathbf{x}) = \langle s | \mathbf{p}| \sigma
\rangle \cdot \mathbf{E}(\mathbf{x})$, operator
$v^\dagger_\sigma(\mathbf{x})$ creates a hole with the spin state $\sigma$
at point $\mathbf{x}$ and $c^\dagger_s(\mathbf{x})$ is the respective
electron creating operator. The equations of motion for the interband
polarizations $p_{\sigma,s}(\mathbf{x}_1,\mathbf{x}_2) =
\aver{c_{s}(\mathbf{x}_2)v_\sigma(\mathbf{x}_1)}$, the electron-electron
$e_{s_1,s_2}(\mathbf{x}_1,\mathbf{x}_2) =
\aver{c^\dagger_{s_1}(\mathbf{x}_1)c_{s_2}(\mathbf{x}_2)}$ and the
hole-hole $h_{\sigma_1,\sigma_2}(\mathbf{x}_1,\mathbf{x}_2) =
\aver{v^\dagger_{\sigma_1}(\mathbf{x}_1)v_{\sigma_2}(\mathbf{x}_2)}$ are
derived from the equation of motion $d\aver{f}/dt = i\aver{[H,f]}$ using
the Hartree-Fock approximation for the terms describing the Coulomb
interaction.

Straightforward calculations give immediately quite cumbersome system of
equations. We provide explicitly only one equation, which shows the
structure of the non-linear Coulomb terms
\begin{equation}\label{eq:SBE_FULL}
  \begin{split}
  i \dot{p}_{\sigma,s}(\mathbf{x}_1,\mathbf{x}_2)  & =
  {K}_{\sigma,s}[p] +
  \left[U_\Delta(\mathbf{x}_1) - U_\Delta(\mathbf{x}_2)\right]{p}_{\sigma,s}(\mathbf{x}_1,\mathbf{x}_2) \\
  & -\int d\mathbf{x}'\,\left[V(\mathbf{x}_1 -\mathbf{x}') - V(\mathbf{x}_2 -\mathbf{x}')\right]
  \left[h_{\sigma',\sigma}(\mathbf{x}',\mathbf{x}_1)p_{\sigma',s}(\mathbf{x}',\mathbf{x}_2)
  - p_{\sigma,s'}(\mathbf{x}_1,\mathbf{x}')e_{s',s}(\mathbf{x}',\mathbf{x}_2)\right] \\
  & + \mathcal{E}_{\sigma,s}(\mathbf{x}_1)\delta(\mathbf{x}_1 - \mathbf{x}_2)
  - h_{\sigma',\sigma}(\mathbf{x}_2,\mathbf{x}_1)\mathcal{E}_{\sigma',s}(\mathbf{x}_2)
  - \mathcal{E}_{\sigma,s'}(\mathbf{x}_1) e_{s',s}(\mathbf{x}_1,\mathbf{x}_2).
  \end{split}
\end{equation}
Here and below a summation over dashed spin indices is implied. This
equation clearly demonstrates that at the diagonal $\mathbf{x}_1 =
\mathbf{x}_2$ the contributions of the Coulomb terms cancel each other. In
Eq.~\eqref{eq:SBE_FULL} $U_\Delta(\mathbf{x})$ describes an effective
background potential created by the local imbalance between the electrons
and holes
\begin{equation}\label{eq:U_Delta_def}
  U_\Delta(\mathbf{x}) = \int d\mathbf{x}'\,V(\mathbf{x}-\mathbf{x}')
  \left[e_{s',s'}(\mathbf{x}',\mathbf{x}') - h_{\sigma',\sigma'}(\mathbf{x}',\mathbf{x}')\right].
\end{equation}
For a semiconductor excited by a single plane wave this potential vanishes
in an overall neutral system due to the translational invariance. In the
two wave excitation setup the effect of $U_\Delta$ is small for bound
exciton states if the order of the multi-wave mixing response is not too
high.

The dynamical equations significantly simplify after the terms being
rearranged to form renormalizations of the single-particle energies and
the coupling between the charge densities and the interband polarizations
\begin{equation}\label{eq:SBE_coordinate}
  \begin{split}
  i \dot{p}_{\sigma,s}(\mathbf{x}_1,\mathbf{x}_2) = & \,
  \widehat{K}_{\sigma,s}[p] + \mathcal{E}_{\sigma,s}(\mathbf{x}_1)\delta(\mathbf{x}_1 - \mathbf{x}_2)
  - \int d\mathbf{x}' \left[
  \widetilde{\mathcal{E}}_{\sigma',s}(\mathbf{x}',\mathbf{x}_2)h_{\sigma',\sigma}(\mathbf{x}',\mathbf{x}_1)
  - \widetilde{\mathcal{E}}_{\sigma,s'}(\mathbf{x}',\mathbf{x}_1)e_{s',s}(\mathbf{x}',\mathbf{x}_2)
  \right] , \\
 -i \dot{e}_{s_1,s_2}(\mathbf{x}_1,\mathbf{x}_2) = & \,
 \widehat{K}_{s_1,s_2}[e] + \int d\mathbf{x}' \left[
  \widetilde{\mathcal{E}}^*_{\sigma',s_1}(\mathbf{x}',\mathbf{x}_1)p_{\sigma',s_2}(\mathbf{x}',\mathbf{x}_2)
  -
  \widetilde{\mathcal{E}}_{\sigma',s_2}(\mathbf{x}',\mathbf{x}_2)p^*_{\sigma',s_1}(\mathbf{x}',\mathbf{x}_1)
  \right], \\
 -i \dot{h}_{\sigma_1,\sigma_2}(\mathbf{x}_1,\mathbf{x}_2) = &\,
 \widehat{K}_{\sigma_1,\sigma_2}[h] + \int d\mathbf{x}' \left[
  \widetilde{\mathcal{E}}^*_{\sigma_1,s'}(\mathbf{x}_1,\mathbf{x}')p_{\sigma_2,s'}(\mathbf{x}_2,\mathbf{x}')
  -
  \widetilde{\mathcal{E}}_{\sigma_2,s'}(\mathbf{x}_2,\mathbf{x}')p^*_{\sigma_1,s'}(\mathbf{x}_1,\mathbf{x}')
    \right],
  \end{split}
\end{equation}
where we have introduced the modified coupling between the interband
polarization and the charge densities
\begin{equation}\label{eq:modified_coupling}
 \widetilde{\mathcal{E}}_{\sigma,s}(\mathbf{x},\mathbf{y}) = \mathcal{E}_{\sigma,s}(\mathbf{x})
 \delta(\mathbf{x}-\mathbf{y}) -
 V(\mathbf{x}-\mathbf{y})p_{\sigma,s}(\mathbf{x},\mathbf{y}).
\end{equation}
We represent the time dependence of the external field in the form
$E(t)=\widetilde{E}(t) e^{-i \Omega t}$, where the amplitude
$\widetilde{E}(t)$ is assumed to be constant during the excitation. In the
rotating frame the operators
\begin{equation}\label{eq:dynamical_operators}
  \begin{split}
  \widehat{K}_{\sigma,s}[p] = & \,
  \widehat{H}_{\sigma,\sigma'}p_{\sigma',s}
   + p_{\sigma,s'}\widehat{H}_{s',s} - V(\mathbf{x}_1-\mathbf{x}_2)p_{\sigma,s}(\mathbf{x}_1,\mathbf{x}_2)
    - \Omega p_{\sigma,s}(\mathbf{x}_1,\mathbf{x}_2), \\
  \widehat{K}_{s_1,s_2}[e] = & \,
  \widehat{H}_{s_1,s'}e_{s',s_2} -
  e_{s_1,s'}\widehat{H}_{s',s_2}, \\
  \widehat{K}_{\sigma_1,\sigma_2}[h] = & \,
  \widehat{H}_{\sigma_1,\sigma'}h_{\sigma',\sigma_2} -
  h_{\sigma_1,\sigma'}\widehat{H}_{\sigma',\sigma_2}
  \end{split}
\end{equation}
are expressed in terms of the integro-differential operators
$\widehat{H}$. The action of these operators is defined by
\begin{equation}\label{eq:integro_diff_SBE}
  \begin{split}
  \widehat{H}_{s_1,s'}f_{s'}=  & \left[-\frac{1}{2m_{s_1}} \frac{\partial^2}{\partial \mathbf{x}_1^2}
    +V_e(\mathbf{x}_1) + U_\Delta(\mathbf{x}_1) + H_{s_1}\right]f_{s_1}(\mathbf{x}_1,\mathbf{x}_2)
    -\int d\mathbf{x}'\, V(\mathbf{x}_1 - \mathbf{x}')e_{s_1,s'}(\mathbf{x}_1,\mathbf{x}')
    f_{s'}(\mathbf{x}',\mathbf{x}_2), \\
  \widehat{H}_{\sigma_1,\sigma'}f_{\sigma'}= &
  -\frac{1}{2m_{\sigma_1,\sigma'}} \frac{\partial^2}{\partial \mathbf{x}_1^2} f_{\sigma'}(\mathbf{x}_1, \mathbf{x}_2)
    +\left[V_h(\mathbf{x}_1) - U_\Delta(\mathbf{x}_1) + H_{\sigma_1}\right]f_{\sigma_1}(\mathbf{x}_1,\mathbf{x}_2) \\
    & \qquad -\int d\mathbf{x}'\, V(\mathbf{x}_1 - \mathbf{x}')h_{\sigma_1,\sigma'}(\mathbf{x}_1,\mathbf{x}')
    f_{\sigma'}(\mathbf{x}',\mathbf{x}_2)
  \end{split}
\end{equation}
with $V_e$ and $V_h$ being the electron and hole confinement potentials,
which effect on the spatial motion of the electrons and holes is estimated
by the $\delta$-functional approximation for the quantum well. The
parameters $H_s$ and $H_\sigma$ denote the energy offsets of the
respective bands. The diagonal elements of $m_{\sigma_1,\sigma_2}$ are the
masses of the light- and heavy-holes, while the off-diagonal elements
account for the valence bands mixing.

Thus, $\widehat{K}_{\sigma,s}$ has the meaning of a two-particle
Hamiltonian with the Coulomb interaction between them, while the
properties of free particles are described by the Hamiltonians
$\widehat{H}_{\sigma_1,\sigma_2}$ and $\widehat{H}_{s_1,s_2}$.


\end{document}